\documentclass[journal,10pt]{IEEEtran}
\ifCLASSINFOpdf
\else
\fi
\usepackage{amsmath}
\usepackage{array}

\usepackage{amsfonts}
\usepackage{amsthm}
\usepackage{amssymb}
\usepackage{graphicx}
\usepackage{caption}
\usepackage[justification=centering]{caption}
\usepackage{booktabs}
\usepackage{arydshln}
\usepackage{multirow}

\usepackage{nomencl}
\usepackage{ifthen}

\renewcommand{\nomgroup}[1]{%
  \item[\textbf{%
    \ifthenelse{\equal{#1}{A}}{Parameters}{}%
    \ifthenelse{\equal{#1}{F}}{Functions}{}%
    \ifthenelse{\equal{#1}{S}}{Sets}{}%
    \ifthenelse{\equal{#1}{V}}{Variables}{}
    }]%
}
\makenomenclature

\usepackage{algorithmic}
\usepackage{fixltx2e}
\usepackage{dblfloatfix}
\begin{document}
%
\title{Two-Stage Submodular Optimization of Dynamic Thermal Rating for Risk Mitigation Considering Placement and Operation Schedule}
%
%
%

\author{Qinfei~Long,~\IEEEmembership{Student~Member,~IEEE,}
        Junhong~Liu,~\IEEEmembership{Student~Member,~IEEE,}
        Chenhao~Ren,~\IEEEmembership{Student~Member,~IEEE,}
        Wenqian~Yin,~\IEEEmembership{Student~Member,~IEEE,}
        Feng~Liu,~\IEEEmembership{Senior~Member,~IEEE,}
        and~Yunhe~Hou,~\IEEEmembership{Senior~Member,~IEEE}
\thanks{This work was supported in part by the National Natural Science Foundation of China (NSFC) under Grant 52177118, and in part by the Research Grants Council of Hong Kong under Grant GRF 17209419.}
\thanks{Qinfei Long, Junhong Liu, Chenhao Ren, Wenqian Yin and Yunhe Hou are with the Department of
Electrical and Electronic Engineering, The University of Hong Kong, Hong
Kong (e-mail: \{qflong, jhliu, chren, wqyin, yhhou\}@eee.hku.hk).}
\thanks{Feng Liu is with the State Key Laboratory of Power Systems, Department of Electrical Engineering, Tsinghua University, Beijing, 100084, China (e-mail: lfeng@tsinghua.edu.cn).}
}

%
%

\markboth{Journal of \LaTeX \ class files}%
{Shell \MakeLowercase{\textit{et al.}}: Bare Demo of IEEEtran.cls for IEEE Journals}
%



\maketitle

\begin{abstract}

Cascading failure causes a major risk to society currently. To effectively mitigate the risk, dynamic thermal rating (DTR) technique can be applied as a cost-effective strategy to exploit potential transmission capability. From the perspectives of service life and Braess paradox, it is important and challenging to jointly optimize the DTR placement and operation schedule for changing system state, which is a two-stage combinatorial problem with only discrete variables, suffering from no approximation guarantee and dimension curse only based on traditional models. Thus, the present work proposes a novel two-stage submodular optimization (TSSO) of DTR for risk mitigation considering placement and operation schedule. Specifically, it optimizes DTR placement with proper redundancy in first stage, and then determines the corresponding DTR operation for each system state in second stage. Under the condition of the Markov and submodular features in sub-function of risk mitigation, the submodularity of total objective function of TSSO can be proven for the first time. Based on this, a state-of-the-art efficient solving algorithm is developed that can provide a better approximation guarantee than previous studies by coordinating the separate curvature and error form. The performance of the proposed algorithms is verified by case results.
\end{abstract}


%
\IEEEpeerreviewmaketitle

\vspace{-10pt}

\section{Introduction}
%
%
%
%

\IEEEPARstart{W}{ith} the fast increase in load demand and integration of renewable energy, power system faces a growing risk of cascading failure (CF), inflicting significant harm to human life and industrial generation \cite{sun2019power}. In 2021, for example, Texas blackout caused by a freezing disaster left more than 10 million people without power and cost more than \$130 billion in industrial damages \cite{busby2021cascading}. Thus, mitigating the CF risk becomes a top priority for each country. 

Conventional CF risk mitigation strategies involve unit commitment \cite{peng2017risk}, topology switching \cite{wang2019markov}, line hardening \cite{liu2018mitigating} and so on. While traditional strategies are beneficial to risk mitigation plans, they commonly need huge investments or are heavily affected by natural factors such as terrain. Luckily, the development of smart sensor enables power system to utilize its potential capability to mitigate risk with a cheaper expenditure. One of the promising technologies is dynamic thermal rating (DTR), which is often placed in overhead lines to dynamically set the thermal rating depending on real-time environment to maximize the usage of existing transmission assets \cite{iglesias2014guide}.

As a result of benefits that DTR can provide, several scientists have analyzed DTR through simulation modeling \cite{zhan2016time}, capacity control \cite{bucher2015robust}, operation scheduling \cite{qiu2014distributionally}, and risk mitigation \cite{wang2018contingency}\cite{xiao2018power}, in which researchers typically assume that DTRs are already placed on each line. One significant drawback is that, due to the vast number of transmission lines, placing DTRs on all lines or the most of them still results in large costs. Therefore, it is critical to place the DTR in the precisely selected lines to maximize its influence.

Sensor placement is a classic combinatorial optimization problem with exponential computational complexity. Assessment indexes \cite{chu1992selecting}, mixed integer programming \cite{zhan2018stochastic}, traditional approximation method \cite{noel2008optimal} and heuristic algorithms \cite{jiang2016novel} have been used to solve placement optimization. However, due to dimensionality curse, some of them are not appropriate for large-scale case, and there is an absence of analytical approximation guarantee, defined as the ratio of the worst obtained sub-optimal value to the real optimal value, potentially resulting in severe performance loss.

Another gap is that fewer research considers the DTR service life \cite{akyildiz2002wireless} following placement. Generally, sensor power sources are limited and generally irreplaceable. While some DTR sensors are designed to be self-powered by feeding off the electromagnetic field \cite{PowerDount}\cite{OverPower}, they are nevertheless difficult to operate throughout the day due to their inefficient charging rate and immature self-powered technique \cite{black2014key}. Also, DTR is prone to failure owing to hostile environments, as DTR in operation is susceptible to high electric and magnetic fields \cite{yang2007survey}. Moreover, excessive operational DTRs may inversely result in higher system risk, named by the Braess paradox \cite{frank1981braess}\cite{long2021analyzing}. Therefore, in addition to planning the DTR placement, its operation schedule also plays a crucial role in achieving a tradeoff between efficient risk mitigation and longer service time.

Furthermore, the problem consisting of placement and operation schedule can be modeled as two-stage optimization, as shown in PMU deployment \cite{sodhi2010optimal}, DG placement \cite{wang2014robust}, fault current limiters stationing \cite{yang2017placement}, charging station planning \cite{deb2021robust} and edge service configuration \cite{nguyen2021two}. Some \cite{wang2014robust}\cite{yang2017placement}\cite{nguyen2021two} contain both discrete and continuous variables, while others \cite{sodhi2010optimal}\cite{deb2021robust} exclusively include the discrete. In this study, DTR placement and operation schedule only comprise the discrete variables in two stages, known as two-stage combinatorial optimization \cite{balkanski2016learning}, which also suffers from similar issues as stated in the sensor placement problem.

To fill the gaps in analytical guarantee and exponential computational complexity, several researchers currently combine two-stage combinatorial optimization with submodular function to form a new research direction, two-stage submodular optimization (TSSO) \cite{balkanski2016learning}\cite{stan2017probabilistic}. It can provide an analytical approximation guarantee from certain solving algorithms, and generate solutions in polynomial time. Specifically, \cite{balkanski2016learning} initially proposed the TSSO for the problem of learning sparse combinatorial representations. After that, \cite{stan2017probabilistic} specializes the TSSO application in the contents of unknown distribution. And TSSO is also applied to streaming scenario \cite{mitrovic2018data}, difference type \cite{liu2020two} and so on.

To investigate the benefit of DTR in mitigating CF risk, this study proposes a two-stage submodular optimization model of DTR’s placement and operation schedule. Based on Markov probability and sampling weight, a submodular sub-function about CF risk mitigation has been built considering Braess paradox. Using the Markov feature, we analytically show that the total objective function of TSSO is of submodularity. Then, by coordinating the separate curvature and related error, a state-of-the-art solving algorithm is designed, giving a superior guarantee larger than current TSSO related guarantees, and obtaining solution in a polynomial time. The contributions of this paper include:

1) We utilize TSSO to establish a DTR model consisting of both placement and operation schedule, which is a two-stage combinatorial optimization with only discrete variable. To the best of the authors’ knowledge, this is the first time that TSSO model is applied to electrical field.

2) We prove that, unlike the existing TSSO analyses, the total objective function of TSSO is submodular when using the Markov feature. Note that this conclusion holds true for any TSSO model that meets the condition: the sub-function has submodular and Markov features.

3) Given TSSO submodularity, we devise an state-of-the-art solving algorithm based on separate curvature, enabling us to analytically derive the performance guarantee, which is better than the current TSSO related guarantees known so far \cite{balkanski2016learning}\cite{stan2017probabilistic}\cite{mitrovic2018data}\cite{li2021two}\cite{yang2021constrained} and traditional guarantee $1-1/e$ (from single stage optimization) \cite{krause2014submodular}\cite{fujishige2005submodular}. This algorithm also lowers the computational complexity of combinatorial optimization from exponential to polynomial.

4) Case studies present the impacts of environment and separate curvature on DTR risk mitigation. Additionally, the comparison of TSSO and one-stage optimization is conducted, indicating that two-stage model can offer a superior mitigation effect for each system state, and prolong the DTR service life. A performance comparison of alternative two-stage strategies is also performed, showing that the suggested method outperforms other strategies.

\vspace{-2pt}

\section{Preliminaries}
\subsection{Dynamic thermal rating and sensor impacted factors}
DTR is an advanced technique that measures the conductor ambient environment and line status, and then transmits data to system operator to dynamically determine a new thermal rating \cite{iglesias2014guide}. DTR is treated as a cost-effective alternative to other traditional hardening strategies, which exploits potential transmission capability while deferring the costly line construction. According to relevant research both in academic \cite{zhan2018stochastic}\cite{jiang2016novel} and industries \cite{billinton1996time}\cite{puffer2012area}, DTR system can boost the current line capacity by 10\%-30\%, with the most increase of 50\% in windy areas. In this study, the goal is to mitigate the CF risk through DTR placement and operation schedule, which is a type of long-term planning issue. Because the effect of DTR on planning issue is evaluated from historical long-term weather data, especially average weather data over a certain period \cite{xiao2018power}\cite{chu1992selecting}\cite{zhan2018stochastic}\cite{jiang2016novel}.

In power system protection, the DTR value can be regarded as a new estimated threshold for relay action. Specifically, this threshold value is predefined to guide the relay action based on data for a period, long enough to ignore the non-steady state in heat balance, allowing the static heat-balance equation to be applied, as described below \cite{xiao2018power}\cite{chu1992selecting}\cite{zhan2018stochastic}\cite{jiang2016novel}:
 
\vspace{-15pt}

\begin{small}
\begin{align} \label{qua:static}
 Q_c + Q_r = {I^2}R + Q_s
\end{align}
\end{small}

\vspace{-5pt}

\noindent where $I^2 R$ denotes the conductor heating from power flow, $Q_s$ the solar radiation heating, $Q_c$ the wind cooling, $Q_r$ the radiation cooling. Specifically, the values of $Q_s$, $Q_c$ and $Q_r$ are connected to the ambient weather condition, and their detailed formulas can be found in \cite{iglesias2014guide}. For convenience, we adopt $I^2 R$, the power as DTR value in the following risk mitigation model.

On the other hand, DTR sensors consist of two different configurations \cite{black2014key}\cite{murphy2013implementation}. The first is located within the substation, charged via a battery system and already connected to data network. The second is used at tower sites, where it is powered by the outside environment and relies only on wireless communications. The tower-based DTR takes the majority in transmission grid, although it has some drawbacks. First, it is hard for sensors to charge themselves directly from high voltage power. Because of immature self-powered technique, it is impractical to keep DTR operating at all time \cite{PowerDount}\cite{OverPower}. To conserve energy, for instance, the DTRs in North Wales system only transmit data at thirty-minute intervals \cite{murphy2013implementation}. Second, since the memory capacity of DTR sensor is limited to few megabytes, the frequent and needless operations will lower the efficiency and accuracy of data communication \cite{akyildiz2002wireless}\cite{yang2007survey}. Finally, the DTR tiny units are subjected to intense electric and magnetic fields as well as strong electric transients induced by switching surges and lightning \cite{yang2007survey}. NERC reported in 2010 that 941 of 3519 tested DTR sensors had discrepancies due to device issues \cite{black2014key}. Thus, excessive operation will accelerate sensor outage, and DTR service life can be extended by having sensors only run for specific periods of time.

\vspace{-15pt}

\subsection{Two-stage Submodular optimization}

Submodularity is defined as the diminishing return in some special set functions. Specifically, it indicates that the incremental “value” of adding components to a set $S_{\rm{A}}$ declines as $S_{\rm{A}}$ is considered grows larger. This property appears in many domains, including economics, computer science, network analysis and so on \cite{krause2014submodular}. It is defined as follows.

\noindent \textbf{Definition 1.} (\cite{fujishige2005submodular}) Suppose set $S_{\rm{A}} \subseteq S_{\rm{B}} \subseteq S_{\rm{L}}$ and set $v\in S_{\rm{L}} \backslash S_{\rm{B}}$, where $S_{\rm{L}}$ is the ground set. A function ${f}:{2^{|{S_L}|}} \to \mathbb{R}$ is submodular if it satisfies
\vspace{-15pt}

\begin{small}
\begin{align}
f(S_{\rm{A}} \cup v) - f(S_{\rm{A}}) \ge f(S_{\rm{B}} \cup v) - f(S_{\rm{B}})
\end{align}
\end{small}

\vspace{-5pt}

In terms of definition, we can obtain the following Lemmas \cite{fujishige2005submodular} which provide a facilitated tool for determining if a function is submodular.

\noindent \textbf{Lemma 1.} If $\alpha_i\geq0$ and ${f_i}:{2^{|{S_L}|}} \to \mathbb{R}$ is submodular function, then so is $\sum_{i}{\alpha_if_i}$.

\noindent \textbf{Lemma 2.} Any submodular function $f$ can be represented as a sum of submodular functions $\widetilde{f}$ and a modular function $m$, i.e., $f=\widetilde{f}+m$.

Based on the submodularity definition, two-stage submodular optimization (TSSO) can be constructed. \cite{balkanski2016learning} initially proposed the TSSO in 2016, i.e., from a selective optimal subset of the ground set in first stage, making the appropriate strategy for each submodular sub-function in second stage. TSSO is one of the types in which the objective functions in two stages are the same or similar, and only contain discrete variables. The detail is shown in $Section$ $IV$. 
According to \cite{balkanski2016learning}, TSSO is a combinatorial counterpart of representation learning tasks, a type of two-stage combinatorial optimization exemplified in dictionary learning \cite{mairal2009online}, autoencoder design \cite{vincent2010stacked} and so on. Following \cite{balkanski2016learning}, \cite{stan2017probabilistic} specializes TSSO application range for unknown distribution, and \cite{mitrovic2018data} extends TSSO to distributed and streaming cases. Furthermore, some studies extend TSSO model. \cite{liu2020two} relaxes TSSO to the problems that are neither non-negative nor monotone. \cite{li2021two} integrates TSSO with the curvature index. Other references also analyze TSSO from different points, such as subsampling model \cite{harshaw2022power}, meta-learning \cite{adibi2020submodular}, extended $P$-matroid constraint \cite{yang2021constrained}. In short, the research of TSSO is still in its infancy, and there are still numerous improvements and application scenarios to be explored.

\vspace{-15pt}

\subsection{Curvature}
The notion of curvature reflects how much a set function’s marginal values might fall. The definition of curvature is below.

\noindent \textbf{Definition 2.} (\cite{conforti1984submodular}) Suppose $S_{\rm{L}}$ is ground set, $f$ a monotone submodular function, and $j \in S_{\rm{L}}$ a single component. The curvature can be defined as

\vspace{-15pt}

\begin{small}
\begin{align}
{\kappa _f} = 1 - \mathop {\min }\limits_{j \in {S_{\rm{L}}}} \frac{{f({S_{\rm{L}}}) - f({S_{\rm{L}}}\backslash j)}}{{f(j) - f(\emptyset )}}
\end{align}
\end{small}

\vspace{-5pt}

 In \cite{sviridenko2017optimal}, the curvature was originally applied to deduce an improved approximation bound, shown in the below lemma.

\noindent \textbf{Lemma 3.} (\cite{sviridenko2017optimal}) If ${f_i}:{2^{|{S_L}|}} \to \mathbb{R}$ is a monotone increasing submodular function with curvature $\kappa_f$, then for all $T \subseteq S_{\rm{L}}$ it has

\vspace{-10pt}

\begin{small}
\begin{align}
\sum\limits_{t \in T} {f(t|{S_{\rm{L}}}\backslash t) \ge (1 - {\kappa _f})f(T)} 
\end{align}
\end{small}

\noindent where $f(t|{S_{\rm{L}}}\backslash t) = f(S_{\rm{L}}) - f({S_{\rm{L}}}\backslash t)$

\section{Dynamic Thermal Rating-based Risk Mitigation Model}
\subsection{Modeling of risk mitigation}
Define $M^l=\{( M^{l(1)},M^{l(2)},\ldots,M^{l(k)},\ldots)$ as the failure chain sub-database from $l$-th system state contained in total database $M$, and $M^{l(k)}=\{({FG}_0^{l(k)},{FG}_1^{l(k)},\ldots,{FG}_i^{l(k)},\ldots)$ as the $k$-th cascading failure chain in $M^l$, where ${FG}_i^{l(k)} \in 2^{|S_{\rm{L}}|}$ denotes the system line state of $i$-th failure generation in $M^{l(k)}$, and $S_{\rm{L}}$ is the line set.

Owing to the piecewise structure, the single component failure probability $\varphi_e({FG}_i^{l(k)})$ is approximated smoothly by Sigmoid form\cite{long2022submodular}:

\vspace{-15pt}

\begin{small}
\begin{align}
{\varphi _e}(FG_i^{l(k)}) = {\Pr}^{\min}_{e}  + \frac{{\Pr_e^{\max }  - \Pr _e^{\min } }}{{1 + \exp [ - \mu \frac{{2P_e - \alpha(P_e^{\min } + P_e^{\max } )}}{{\alpha (P_e^{\min } + P_e^{\max }) }}]}}
\end{align}
\end{small}

\vspace{-5pt}

\noindent where $P_e$, $P_e^{\min}$, $P_e^{\max}$ respectively denote current power flow, minimum and maximum transmission capabilities of line $e$, $\Pr_e^{\min}$, $\Pr_e^{\max}$ the minimum and maximum of failure probability, respectively, and $\mu$ the approximate factor to avoid the vanishing gradient problem. And $\alpha$ denotes the DTR improved parameter for power transmission threshold, improving the original to $\alpha P_e^{\max }$ related to ambient weather \cite{iglesias2014guide}. When $\alpha=1$ there is no DTR effect.

For cascading failure chain with Markovian property \cite{guo2017toward}, we define ${fp}_{e}^{l(k)}$ as the probability of a specific cascading failure chain $M^{l(k)}$ with DTR placed in line $e$:

\vspace{-15pt}

\begin{small}
\begin{align}
 {fp}_{e}^{l(k)} & = \prod\limits_{i = 1}^d {[\prod\limits_{j \in \overline {S_{_{{fp_i}}}^{{\rm{non}}}} } {{\varphi _j}(FG_i^{l(k)})} }  \cdot \prod\limits_{j \in S_{_{{fp_i}}}^{{\rm{non}}}} {(1 - {\varphi _j}(FG_i^{l(k)})} )] \notag\\
 & \quad ... \cdot {\rm H^ {\textit{l}(\textit{k})}}(\varphi {'_e}) 
\end{align}
\end{small}

\vspace{-5pt}

\noindent where $\overline {S_{_{{fp_i}}}^{{\rm{non}}}}$ denotes the no-DTR line set that functioned normal in ${FG}_{i-1}^{l(k)}$ but failed in ${FG}_{i}^{l(k)}$, $S_{_{{fp_i}}}^{{\rm{non}}}$ denotes the no-DTR line set functioned normal in ${FG}_i^{l(k)}$, and $\rm{H}^{\textit{l}(\textit{k})}({\varphi\prime}_e)$ is defined as

\vspace{-15pt}

\begin{small}
\begin{align} \label{func:H^k'}
{\rm H^ {\textit{l}(\textit{k})}}(\varphi {'_e}) = \left\{ {\begin{array}{*{20}{c}}
   {\prod\limits_{i = 1}^d {(1 - \varphi {'_e}(FG_i^{l(k)}))} } & {d_e > d}  \\
   {\varphi {'_e}(FG_{d_e}^{l(k)}) \cdot \prod\limits_{i = 1}^{d_e - 1} {(1 - \varphi {'_e}(FG_i^{l(k)}))} } & {{\rm{Other}}}  \\
\end{array}} \right.
\end{align}
\end{small}

\vspace{-10pt}

\noindent where $d_e$ is the generation line $e$ fails. If $d_e >d$, it means there is no failure in line $e$ during $M^{l(k)}$. For ${fp}_{\rm{B}}^{l(k)}$ in which lines set $S_{\rm{B}}$ placed with DTR, it is similar to the above, whose details can be found in our previous research\cite{long2022submodular}.

Moreover, define load loss induced by a single failure chain $M^{l(k)}$ as $Y_{l(k)}$ from cascading failure simulator \cite{long2021analyzing}. Since cascading failure is treated as an infrequent but extreme event, which has less occurrence probability but can bring great damage to the whole power system, we define $\delta_{\{Y_{l(k)}>Y_{ext}\}}$ as an indicator that selects the cascading failure chain in which the load loss exceeds $Y_{ext}$, the setting loss threshold. Specifically, $\delta_{\{Y_{l(k)}>Y_{ext}\}}=1$ when $Y_{l(k)}>Y_{ext}$, otherwise $\delta_{\{Y_{l(k)}>Y_{ext}\}}=0$. 

Then we propose the original risk of a single failure chain $M^{l(k)}$ as

\vspace{-17pt}

\begin{small}
\begin{align}
Risk_\emptyset ^{l(k)} = Y_{l(k)} \cdot {fp_\emptyset^{l(k)}} \cdot {\delta _{\{ {Y_{l(k)}} > Y_{ext}\} }}
\end{align} 
\end{small}

\vspace{-5pt}

Above all, the original risk of whole failure chains in $M^l$ can be modeled as

\vspace{-17pt}

\begin{small}
\begin{align} \label{fun:riskw_null}
  RiskW_\emptyset^{l} & = \mathbb{E} (Y \cdot {\delta _{\{ Y > Y_{ext}\} }}) \notag\\ 
 & = \sum\limits_{{M^{l(k)}} \in M^{l}} {Risk_\emptyset ^{l(k)}}  \\ 
 & = \frac{1}{{|M^l|}}\sum\limits_{{M^{l(k)}} \in M^{l}} {Y_{l(k)} \cdot {\delta _{\{ {Y_{l(k)}} > Y_{ext}\} }}}  \notag  
\end{align}
\end{small}

Similarly, if DTR is placed in set $S_{\rm{B}}$, there are $Risk_{\rm{B}} ^{l(k)}$ and $RiskW_{\rm{B}}^{l}$ \cite{long2022submodular}.

\vspace{-15pt}

\subsection{Sampling weight technique}
When some parameters change, such as line maximum capacity after DTR placement, the related cascading failure chains may be changed, impacting failure chain probability and risk index. However, recreating the failure database based on the new system will impose a considerable computing burden, particularly in optimization issues. To resolve this problem, the sampling weight technique proposed in \cite{guo2017toward} is applied.

\subsubsection{Sampling weight in a single failure chain}
In $M^{l(k)}$, given 2 different line sets $S_{\rm{A}}$ and $S_{\rm{B}}$ placed with DTR, the underlying relationship between ${fp}_{\rm{A}}^{l(k)}$ and ${fp}_{\rm{B}}^{l(k)}$ can be expressed as sampling weight

\vspace{-15pt}

\begin{small}
\begin{align} \label{func:W_(A-B)1}
W_{{\rm{A}} - {\rm{B}}}^{l(k)} \! = \! \frac{{fp_{\rm{A}}^{l(k)}}}{{fp_{\rm{B}}^{l(k)}}} \! = \! \prod\limits_{e \in {S_{\rm{A}}}\backslash {S_{\rm{B}}}} \! {\frac{{{{\rm H}^{l(k)}}(\varphi {'_{{\rm{A}}e}})}}{{{{\rm H}^{l(k)}}({\varphi _{{\rm{B}}e}})}}} \! \cdot \! \prod\limits_{e \in {S_{\rm{B}}}\backslash {S_{\rm{A}}}} \! {\frac{{{{\rm H}^{l(k)}}({\varphi _{{\rm{A}}e}})}}{{{{\rm H}^{l(k)}}(\varphi {'_{{\rm{B}}e}})}}}
\end{align}
\end{small}

\vspace{-5pt}

\noindent where $\varphi {'_{{\rm{A}}e}}$ and $\varphi {'_{{\rm{B}}e}}$ denote the failure probabilities of line $e$ when DTR placed in $S_{\rm{A}}$ and $S_{\rm{B}}$, respectively, and $\varphi {_{{\rm{A}}e}}$ and $\varphi {_{{\rm{B}}e}}$ the original probabilities. 
Note that $S$ can be empty set. From formula (\ref{func:W_(A-B)1}), if ${S_{\rm{B}}}  \subseteq {S_{\rm{A}}}$ the value of ${fp}_{\rm{A}}^{l(k)}$ can be calculate by multiplying $W_{{\rm{A}} - {\rm{B}}}^{l(k)}$ by ${fp}_{\rm{B}}^{l(k)}$, where the calculation of ${fp}_{\rm{B}}^{l(k)}$ also can be similarly decomposed, showing the Markov feature \cite{long2022submodular}\cite{guo2017toward}.

\subsubsection{General sampling weight for the optimal combination}

Define $G_i (S)=\{T \subseteq S:T=argmax f_i (\cdot)\}$ as the optimal solution for $f_i (\cdot)$ in feasible region $S$. For the facilitation of model, the term $\overset{\frown}{W}$ is used to represent the general sampling weight, the ratio of optimal objective values in different feasible region. Specifically, for sets ${S_{\rm{A}}}$ and ${S_{\rm{B}}}$, we have

\vspace{-15pt}

\begin{small}
\begin{align}
{\mathord{\buildrel{\lower3pt\hbox{$\scriptscriptstyle\frown$}} 
\over W} _{{\rm{B - A}}}} = \frac{{{f_i}({G_i}({S_{\rm{B}}}))}}{{{f_i}({G_i}({S_{\rm{A}}}))}}
\end{align}
\end{small}

\vspace{-10pt}

Similarly, $\overset{\frown}{W}_{\rm{A}-\emptyset}$ and $\overset{\frown}{W}_{\rm{B} \cup v - \rm{A} \cup v}$ can be obtained.

\vspace{-5pt}

\section{TSSO-based risk mitigation optimization}
Since system has different states in terms of load and generation, a unique DTR placement (or operation) scheme cannot provide the best risk mitigation effect in each state. On the contrary, an inappropriate DTR operation may result in Braess paradox, i.e., the improvement or adding of some components worsens the system performance \cite{long2022submodular}, initially discovered in transportation research \cite{frank1981braess}. Additionally, DTR service life is also an essential aspect in planning that has received less attention in the relevant research \cite{zhan2016time}\cite{bucher2015robust}\cite{qiu2014distributionally}\cite{wang2018contingency}\cite{xiao2018power}. A better DTR operation schedule not only helps the system mitigate risk more flexibly and pertinently, also prolongs the DTR service life, lowering the failed rate. Compared to one-stage optimization, the TSSO-based method will place DTR with reasonable redundancy, providing additional options for designing the associated operation schedules. In this section, we will establish a DTR risk mitigation optimization model based on TSSO that consists of placement and operation schedule. 

\vspace{-15pt}

\subsection{Submodular optimization for a single system state}
In order to quantify the DTR placement effect on CF risk mitigation, define ${S_{\rm{B}}} \subseteq S_{\rm{L}}$ as an allocation of potential lines with DTR placement. Then we want to find $S_{\rm{B}}$ to maximize the risk mitigation effect in $M^l$ that

\begin{small}
\begin{align} \label{fun:subobj}
 f_{l}({S_{\rm{B}}}) = RiskW_\emptyset^{l} - RiskW_{\rm{B}}^{l} + \eta \cdot BPI 
\end{align}
\end{small}

\vspace{-8pt}

\noindent where $\eta\in(0,1]$ denotes an adjustment factor,
$BPI=\sum_{M^{l(k)}\in M^l}{[\max{\left(w_{com} ^{l(k)}-1,\ 0\right)} \cdot {Risk}_{\rm{A}}^{l(k)}}]$ denotes Braess paradox indicator, with a higher value indicating more side effects of Braess paradox. 
$w_{com}^{l(k)}=\frac{{Risk}_{\rm{B}}^{l(k)}}{{Risk}_{\rm{A}}^{l(k)}}$ denotes the compared sampling weight between new updating set $S_{\rm{B}}$ with previous updating set $S_{\rm{A}}$. The model detail can be found in our prior work\cite{long2022submodular}.

\noindent \textbf{Theorem 1.}(\cite{long2022submodular}) Sub-function $f_{l}({S_{\rm{B}}})$ is submodular.

\vspace{-10pt}

\subsection{Construction of TSSO}
In two-stage submodular optimization (TSSO) application on risk mitigation model, the target is to select a subset of lines to place DTR in the first stage, and then schedule the DTR operation mode (on/off) in a series of sub-functions to optimize the target value in the second stage. The sub-functions correspond to different system states.

Without loss of generality, let TSSO contains a class of sub-functions $f_1 (\cdot)$,$f_2 (\cdot)$,$\ldots$ in different system states, where sub-function ${f_i}:{2^{|{S_{\rm{L}}}|}} \to \mathbb{R}$ is defined over $S_{\rm{L}}$. The aim is to find a set $S \subseteq S_{\rm{L}}$ of size less than $k_{c1}$, whose subsets $T_1$,$T_2$,$\ldots$,$T_i$,$\ldots$ with sizes less than constraint $k_{c2}$, to maximize the mean value $F(S)$:

\vspace{-19pt}

\begin{small}
\begin{align} \label{fun:TSSOobj}
 \mathop {\max }\limits_{S:|S| \le {k_{c1}}} F(S) = \mathop {\max }\limits_{S:|S| \le {k_{c1}}} \mathbb{E} \mathop {\max }\limits_{{\rm{ }}{T_i} \in S,|{T_i}| \le {k_{c2}}} {f_i}({T_i})   
\end{align}
\end{small}

\vspace{-3pt}

In this study, we can set the sub-function $f_i$ as the risk mitigation function (\ref{fun:subobj}) to build TSSO. Note that even if a class of sub-functions $f_i$ are submodular in second stage, the total objective function $F$ ceases to be submodular in general case, like partition matching \cite{balkanski2016learning}. Because TSSO selects variable $S$ in first stage to serve as the feasible region for second stage optimizations. Based on a new feasible region, the sub-functions optimization may produce a completely different result compared with the result of previous feasible region. In previous studies \cite{balkanski2016learning}\cite{stan2017probabilistic}\cite{mitrovic2018data}\cite{liu2020two}\cite{li2021two}, they all adopt this precondition in TSSO, which greatly restrict the solving performance and have an inferior approximation guarantees. The main contribution of this research is proving that the total function $F$ is submodular when its sub-functions are of Markov and submodular features, as shown in Appendix in detail.

\section{Separate curvature greedy solving algorithm}
The TSSO as formula (\ref{fun:TSSOobj}) is a two-stage combinatorial optimization that basically requires exponential computation complexity to traverses all candidate combinations to obtain the optimal solution, which is a NP-hard problem suffering from dimension curse. To get a sub-optimal solution with an acceptable computing time, it is appropriate to design a solving method with a reduced computation complexity, lowering the original exponential complexity to a polynomial-time one \cite{krause2014submodular}. Previously, the solving algorithms were designed as index-based \cite{long2021analyzing}\cite{xu2020mitigating}, greed-based \cite{balkanski2016learning}\cite{sviridenko2017optimal} or surrogate-based strategies \cite{stan2017probabilistic}\cite{li2021two}\cite{yang2021constrained}. But with the precondition that the total objective function of TSSO is not submodular, they produce a suboptimal result with inferior approximation guarantee. In this part, we propose a state-of-the-art solving algorithm based on separate curvature \cite{conforti1984submodular} to deal with TSSO using the submodularity of TSSO’s total objective function.

\subsection{Solving algorithm design}
In any system state, sub-function $f_i (T_i)$ can be divided into 2 parts, i.e.,

\vspace{-15pt}

\begin{small}
\begin{align}
    {f_i}({T_i}) = {g_i}({T_i}) + {c_i}({T_i})
\end{align}
\end{small}

\vspace{-10pt}

\noindent where $g_i (T_i)$ is a monotone non-negative submodular function. Unlike the typical modular function settings in \cite{liu2020two}\cite{li2021two}\cite{sviridenko2017optimal}, the modular function $c_i (T_i )$ in this work is separate as

\vspace{-15pt}

\begin{footnotesize}
\begin{align}
 {c_i}({T_i}) &= {c_{i1}}({T_i}) + {c_{i2}}({T_i}) \\ 
  &= \sum\limits_{x \in {T_i} \cap {S_{{\rm{L}}1}}} {{f_i}(x|{S_{{\rm{L}}1}}\backslash x)}  + \sum\limits_{x \in {T_i} \cap {S_{{\rm{L}}2}}} {{f_i}(x|{S_{{\rm{L}}2}}\backslash x)}  \notag 
\end{align}
\end{footnotesize}

\vspace{-5pt}

\noindent where $S_{\rm{L}1} \subseteq S_{\rm{L}}$ and $S_{\rm{L}2} = S_{\rm{L}} \backslash S_{\rm{L}1}$. 

Based on the divided form of sub-function, the solving algorithm can be designed. For convenience, the following notations are defined. Let $\Delta _i^g(x,T_i^j)= g_i (x \cup T_i^j) - g_i (T_i^j) $ denotes the marginal increasement of element $x$ adding to set $T_i^j$, $\nabla _i^g(x,y,T_i^j)= g_i (x \cup T_i^j \backslash y)- g_i (T_i^j)$ the marginal gain of replacing $y$ in set $T_i^j$ with component $x \notin T_i^j$, and $i$, $j$ denote the indexes of sub-function and searching iteration, respectively. In addition, there are differences

\vspace{-15pt}

\begin{small}
\begin{align}
    {\Delta _i}(x,T_i^j) = {(1 - \frac{p}{k})^{k - j}}\Delta _i^g(x,T_i^j) + {c_i}(x)
\end{align}
\end{small}

\vspace{-18pt}

\begin{small}
\begin{align}
    {\nabla _i}(x,y,T_i^j) = {(1 - \frac{p}{k})^{k - j}}\nabla _i^g(x,y,T_i^j) + {c_i}(x) - {c_i}(y)
\end{align}
\end{small}

\vspace{-5pt}

\noindent where $k$, $p$ denote the first stage cardinality constraint and constraint’s number of second stage, respectively. Note that $p$ is not the second stage cardinality constraint $k_{c2}$. When component $x$ replaces the component in $T_i^j$, the new set does not violate the constraints polytope defined as $I(S)$.

Above all, the marginal gain of component $x$ in each iteration can be expressed as

\vspace{-15pt}

\begin{footnotesize}
\begin{align}
    {\nabla _i}(x,T_i^j) \! = \! \left\{ \! {\begin{array}{*{20}{c}}
   {{\Delta _i}(x,T_i^j)} \! & \! {if \ T_i^j \cup x \in I(S)}  \\
   {\max \{ 0,\! \mathop {\max }\limits_{y:(T_i^j \cup x)\backslash y \in I(S)} \! {\nabla _i}(x,y,T_i^j)\} } \! & \! {Other}  \\
\end{array}} \right.
\end{align}
\end{footnotesize}

\vspace{-10pt}

Similarly, ${\rm{Re}}{{\rm{p}}_i}(x,T_i^j)$ is defined to represent the element that would be replaced by component $x$ as follows:

\vspace{-15pt}

\begin{footnotesize}
\begin{align}
    {\rm{Re}}{{\rm{p}}_i}(x,T_i^j)\! =\! \left\{ \! {\begin{array}{*{20}{c}}
   \emptyset  \! &  \! {if \ T_i^j \cup x \in I(S)}  \\
   {\mathop {\arg \max }\limits_{y:(T_i^j \cup x)\backslash y \in I(S)} \! {\nabla _i}(x,y,T_i^j)} \! & \! {Other}  
\end{array}} \right.
\end{align}
\end{footnotesize}

\vspace{-10pt}

After the notation definition, the solving algorithm is constructed in Algorithm 1. In detail, this algorithm works in $k=k_{c1}$ rounds, and a carefully designed searching objective function derived from separate curvature is used to select a component $x$ that maximizes the marginal value in each iteration. 

\vspace{-15pt}

\subsection{Performance analysis}
Based upon the designed solving algorithm, the approximation guarantee can be derived. Given the first stage result $S^j$ in each iteration $j=1,…,k$, we define the surrogate function ${\Phi _j}({S^j})$ as

\vspace{-15pt}

\begin{small}
\begin{align}
    {\Phi _j}({S^j}) = \sum\limits_{i = 1}^m {[{{(1 - \frac{p}{k})}^{k - j}}{g_i}(T_i^j) + {c_i}(T_i^j)} ]
\end{align}
\end{small}

\vspace{-5pt}

\noindent where $k$, $p$ denote the first stage cardinality constraint and constraint’s number of second stage, respectively, $m$ is the number of sub-functions, and $T_i^j$, $T_i^*$ are the corresponding searching result and practical optimum result for $i$-th sub-function, respectively. First, there are Lemmas 4 and 5 about the boundary.

\noindent \textbf{Lemma 4.}  For $j=1,2,…,k$, it has
\vspace{-10pt}

\begin{small}
\begin{align}
    {\Phi _j}({S^j}) \!-\! {\Phi _{j - 1}}({S^{j \!-\! 1}}) \!= \!\sum\limits_{i = 1}^m {[{\nabla _i}({x^j},T_i^{j - 1}) \!+\! \frac{p}{k}} {(1 - \frac{p}{k})^{k\! -\! j}}\!{g_i}(T_i^{j\! - \!1})]
\end{align}
\end{small}

\vspace{-15pt}

\noindent \textbf{Proof:} 

\vspace{-15pt}

\begin{small}
\begin{align}
 &\quad {\Phi _j}({S^j}) - {\Phi _{j - 1}}({S^{j - 1}}) \notag\\ 
  &= \sum\limits_{i = 1}^m {[{{(1 - \frac{p}{k})}^{k - j}}{g_i}(T_i^j) + {c_i}(T_i^j)}  - {(1 - \frac{p}{k})^{k - (j - 1)}}{g_i}(T_i^{j - 1}) - {c_i}(T_i^{j - 1})] \notag\\ 
  &=\! \sum\limits_{i = 1}^m \! {[{{(1 \!-\! \frac{p}{k})}^{k\! -\! j}}\!({g_i}(T_i^j) \!-\! {g_i}(T_i^{j \!-\! 1})) \!+\! {c_i}(T_i^j)\! -\! {c_i}(T_i^{j\! - \!1})\! +\! \frac{p}{k}}\! {(1 \!-\! \frac{p}{k})^{k\! -\! j}}\!{g_i}(T_i^{j\! -\! 1})] \notag\\ 
  &= \sum\limits_{i = 1}^m {[{\nabla _i}({x^j},T_i^{j - 1}) + \frac{p}{k}} {(1 - \frac{p}{k})^{k - j}}{g_i}(T_i^{j - 1})] 
\end{align}
\end{small}

\vspace{-10pt}

\noindent where $x^j= \arg {\max _{x \in {S_{\rm{L}}}}}\sum\nolimits_{i = 1}^m {{\nabla _i}(x,T_i^{j - 1})}$.  \hfill $\square$\par

\noindent \textbf{Lemma 5.}  In each iteration of Algorithm 1, if $x^j$ is added to $S^{j-1}$, then

\vspace{-15pt}

\begin{small}
\begin{align}
    \sum\limits_{i = 1}^m \! {{\nabla _i}({x^j}\!,\!T_i^{j - 1})} \! \ge \! \frac{1}{k}\sum\limits_{i = 1}^m {[{{(1\! - \! \frac{p}{k})}^{k \!- \!j}}\!({g_i}(T_i^*) \!-\! p \! \cdot \! {g_i}(T_i^{j - 1})) 
    \!+\! (1 \!-\! O(\xi )){c_i}(T_i^*) ]} 
\end{align}
\end{small}

\vspace{-15pt}

\noindent \textbf{Proof:} 

\vspace{-15pt}

\begin{small}
\begin{align}
  &\quad k\sum\limits_{i = 1}^m {{\nabla _i}({x^j},T_i^{j - 1})}  \notag\\ 
  & \ge k\sum\limits_{i = 1}^m {\max \{ 0,\mathop {\max }\limits_{y:(T_i^j \cup x)\backslash y \in I(S)} {\nabla _i}(x,y,T_i^{j - 1})\} }  \notag\\ 
  & \ge \sum\limits_{i = 1}^m {|T_i^*|\max \{ 0,\mathop {\max }\limits_{y:(T_i^j \cup x)\backslash y \in I(S)} {\nabla _i}(x,y,T_i^{j - 1})\} }  \notag\\ 
  & \ge \sum\limits_{i = 1}^m {|T_i^*|\mathop {\max }\limits_{x \in T_i^*} {\nabla _i}(x,y,T_i^{j - 1})}  \\ 
  & = \sum\limits_{i = 1}^m {|T_i^*|\mathop {\max }\limits_{x \in T_i^*} [{{(1 - \frac{p}{k})}^{k - j}}\nabla _i^g(x,y,T_i^{j - 1}) + {c_i}(x) - {c_i}(y)]}  \notag\\ 
  & \ge \sum\limits_{i = 1}^m {[{{(1 - \frac{p}{k})}^{k - j}}({g_i}(T_i^*) - {g_i}(T_i^{j - 1})) + {c_i}(T_i^*) - {c_i}(T_i^{j - 1})]}  \notag\\ 
  & \ge \sum\limits_{i = 1}^m {[{{(1 - \frac{p}{k})}^{k - j}}({g_i}(T_i^*) - p \cdot {g_i}(T_i^{j - 1})) + (1 - O(\xi )){c_i}(T_i^*)]}  \notag
\end{align}
\end{small}

\vspace{-5pt}

\noindent In above formulation, the first inequality is due to the restricted maximization of ${{\nabla _i}(x,T_i^{j - 1})}$. The second inequality arises from $|T_i^* | \leq k$ for $\forall i$. The third inequality follows since the element $x$ is the best result chosen by algorithm in current searching process compared to components in $T_i^*$. The fifth inequality holds owing to the submodularity of sub-functions. And the final one is correct because $p \geq 1$. Define $O(\xi ) = {c_i}(T_i^{j - 1})/{c_i}(T_i^*)$ as the ratio, in which the difference between $c_i (T_i^{j-1})$ and $c_i (T_i^*)$ is acceptable. Then the prove is over. \hfill $\square$\par

\begin{table}[] \scriptsize
\begin{tabular}{ll}
\hline
\multicolumn{2}{l}{\textbf{Algorithm 1} TSSO risk mitigation solving algorithm (SCG)} \\ \hline
\multicolumn{2}{l}{\begin{tabular}[c]{@{}l@{}}\textbf{Input:} Failure chain database $M$ with $|M|=m$, load loss vector $Y$ and related \\  threshold $Y_{ext}$, line set $S_{\rm{L}}$, line capacity maximum vector ${P_{\max}}$, DTR inproved\\  parameter $\alpha$, sub-functions $g_i (T_i)$ and $c_i (T_i)$, and constraints $k$ and $k_{c2}$.\end{tabular}} \\
\multicolumn{2}{l}{\textbf{Output:} DTR placed set $S_k$ and its operated subsets $T_1 ^k,...,T_m ^k$ in each state.}    \\
1) & $S^0\gets\emptyset$ and $T_i^0\gets \emptyset$ for $i \in [1,m]$    \\
2)   & \textbf{for} $j$=1 to $k$ \textbf{do}                    \\
3)   & \quad    $x^j \gets \arg {\max _{x \in {S_{\rm{L}}}}}\sum\nolimits_{i = 1}^m {{\nabla _i}(x,T_i^{j - 1})} $   \\
4)   & \quad \textbf{if} $\sum\nolimits_{i = 1}^m {{\nabla _i}(x^j,T_i^{j - 1})}>0$ \textbf{then}  \\
5)   & \qquad  $S^j\gets S^{j-1} + x^j$ \\
6)   & \qquad \textbf{for} $i$=1 to $m$ \textbf{do}\\
7)   & \qquad \quad  \textbf{if} ${{\nabla _i}(x^j,T_i^{j - 1})}>0$ \textbf{then} \\
8)   & \qquad \qquad $T_i ^j \gets T_i ^{j-1} \cup x^j \backslash {\rm{Re}}{{\rm{p}}_i}(x^j,T_i^{j-1}) $    \\
9)  & \qquad \quad  \textbf{else do}   \\
10)  & \qquad \qquad $T_i ^j \gets T_i ^{j-1}$ \\
11)  & \qquad \quad \textbf{end if} \\
12)  & \qquad \textbf{end for} \\
13)  & \quad \textbf{end if} \\ 
14)  & \textbf{end for} \\
15)  & Return $S^k$ and $T_1 ^k,...,T_m ^k$ \\ \hline

\end{tabular}%
\vspace{-20pt}
\end{table}

The above lemmas imply the initial approximation ratio of Algorithm 1.

\noindent \textbf{Theorem 2.} Algorithm 1 returns a set $S^k$ of size $k$ such that

\vspace{-15pt}

\begin{small}
\begin{align}
    \sum\limits_{i = 1}^m {[{g_i}(T_i^j)\! +\! {c_i}(T_i^j)} ] \!\ge \!\sum\limits_{i = 1}^m {[\frac{1}{p}(1 - {e^{ - p}})} {g_i}(T_i^*) \!+\! (1 - O(\xi )){c_i}(T_i^*)]
\end{align}
\end{small}

\vspace{-10pt}

\noindent \textbf{Proof:} In accordance with the definition of surrogate function ${\Phi _j}({S^j})$, we have ${\Phi _0}({S^0}) = 0$ and ${\Phi _k}({S^k}) = \sum\limits_{i = 1}^m {[{g_i}(T_i^j) + {c_i}(T_i^j)} ]$. Combining Lemmas 4 and 5, it has ${\Phi _j}({S^j}) - {\Phi _{j - 1}}({S^{j - 1}}) \ge \frac{1}{k}\sum\limits_{i = 1}^m {[{{(1 - \frac{p}{k})}^{k - j}}{g_i}(T_i^*) + (1 - O(\xi )){c_i}(T_i^*)]} $. Thus

\vspace{-10pt}

\begin{scriptsize}
\begin{align}
 &\quad \sum\limits_{i = 1}^m {[{g_i}(T_i^j)+{c_i}(T_i^j)}]\notag\\
 &=\sum\limits_{j=1}^k {[{\Phi_j}({S^j})-{\Phi_{j-1}}({S^{j-1}})}]\notag\\
 &\ge\sum\limits_{j=1}^k {\frac{1}{k}\sum\limits_{i=1}^m {[{{(1-\frac{p}{k})}^{k-j}}{g_i}(T_i^*)+(1-O(\xi)){c_i}(T_i^*)]}}\\
 &=\sum\limits_{i=1}^m {[\frac{1}{p}(1-{{(1-\frac{p}{k})}^k})} {g_i}(T_i^*)+(1-O(\xi)){c_i}(T_i^*)]\notag\\
 &\ge\sum\limits_{i=1}^m {[\frac{1}{p}(1-{e^{-p}})}{g_i}(T_i^*)+(1-O(\xi)){c_i}(T_i^*)]\notag
\end{align}
\end{scriptsize}

\vspace{-8pt}

\noindent In above formulation, the first equality is from definition of ${\Phi _j}({S^j})$ and $f_i (T_i)$. The third equality arises from geometric series sum. The last one holds due to constant inequality ${(1 - \frac{p}{k})^k} \le {e^{ - p}}$. The prove is end.  \hfill $\square$\par

What is more, the relationship between $c_i (T_i )$ and $f_i (T_i )$ is clarified in Lemma 5 based on separate curvature.

\noindent \textbf{Lemma 6.} According to the Definition 2, define the separate curvatures for $f(T \cap {S_{{\rm{L1}}}})$ and $f(T \cap {S_{{\rm{L2}}}})$ as {\small ${\kappa _{f1}}  = 1 - \mathop {\min }\limits_{j \in T \cap {S_{{\rm{L1}}}}}\! \frac{{f({S_{{\rm{L1}}}}) - f({S_{{\rm{L1}}}}\backslash j)}}{{f(j) - f(\emptyset )}}$ } and {\small ${\kappa _{f2}} \!=\! 1\! - \mathop {\min }\limits_{j \in T \cap {S_{{\rm{L2}}}}}\! \frac{{f({S_{{\rm{L2}}}}) - f({S_{{\rm{L2}}}}\backslash j)}}{{f(j) - f(\emptyset )}}$ }. Then there is

\vspace{-14pt}

\begin{small}
\begin{align}
    {c_i}({T_i}) = {c_{i1}}({T_i}) + {c_{i2}}({T_i}) \ge (1 - {\kappa _{f1}} + O({c_2})){f_i}({T_i})
\end{align}
\end{small}

\vspace{-5pt}

\noindent \textbf{Proof:} Due to the submodularity of $f_i (T_i )$ and $S_{\rm{L}2} = S_{\rm{L}} \backslash S_{\rm{L}1}$, we have

\vspace{-15pt}

\begin{small}
\begin{align} \label{fun:lemma5}
    {f_i}({T_i} \cap {S_{{\rm{L}}1}}) + {f_i}({T_i} \cap {S_{{\rm{L}}2}}) \ge {f_i}({T_i})
\end{align}
\end{small}

\vspace{-10pt}

\noindent Then based upon Lemma 3 and (\ref{fun:lemma5}), it has

\vspace{-15pt}

\begin{small}
\begin{align}
 &\quad {c_{i1}}({T_i}) + {c_{i2}}({T_i}) \notag\\ 
  & = \sum\limits_{x \in {T_i} \cap {S_{{\rm{L}}1}}} {{f_i}(x|{S_{{\rm{L}}1}}\backslash x)}  + \sum\limits_{x \in {T_i} \cap {S_{{\rm{L}}2}}} {{f_i}(x|{S_{{\rm{L}}2}}\backslash x)}  \notag\\ 
  & \ge (1 - {\kappa _{f1}}){f_i}({T_i} \cap {S_{{\rm{L}}1}}) + (1 - {\kappa _{f2}}){f_i}({T_i} \cap {S_{{\rm{L}}2}}) \\ 
  & \ge (1 - {\kappa _{f1}})[{f_i}({T_i}) - {f_i}({T_i} \cap {S_{{\rm{L}}2}})] + (1 - {\kappa _{f2}}){f_i}({T_i} \cap {S_{{\rm{L}}2}}) \notag\\ 
  & = (1 - {\kappa _{f1}} + O({c_2})){f_i}({T_i}) \notag 
\end{align}
\end{small}

\vspace{-10pt}

\noindent where $O({c_2}) = ({\kappa _{f1}} - {\kappa _{f2}}){f_i}({T_i} \cap {S_{{\rm{L}}2}})/{f_i}({T_i})$.  \hfill $\square$\par

Finally, the improved approximation ratio of TSSO can be shown in Theorem 3.

\noindent \textbf{Theorem 3.} For $O(\xi )$, $O(c_2 )$ and ${\kappa _{f1}} \in [0,1]$, Algorithm 1 returns a set $S^k$ of size $k$ such that

\vspace{-15pt}

\begin{small}
\begin{align} \label{ieq:TSSOguarantee}
    F({S^k}) \! \ge \! [1 \! - \! \frac{{{\kappa _{f1}}{e^{ - p}}}}{p} \! + \! \frac{{{\kappa _{f1}}}}{p} \! - \! {\kappa _{f1}} \! + \! O(\xi ,{c_2})] \!  F({S^*}) 
\end{align}
\end{small}

\vspace{-5pt}

\noindent \textbf{Proof:} From the Theorem 2 and Lemma 6, we can further obtain that

\vspace{-15pt}

\begin{scriptsize}
\begin{align}  \label{fun:ineq1}
 &\quad \sum\limits_{i = 1}^m {[{g_i}(T_i^j) + {c_i}(T_i^j)} ] \notag\\ 
  & \ge \sum\limits_{i = 1}^m {[\frac{1}{p}(1 - {e^{ - p}})} {g_i}(T_i^*) + (1 - O(\xi )){c_i}(T_i^*)] \notag\\ 
  & = \sum\limits_{i = 1}^m {[\frac{1}{p}(1 - {e^{ - p}})} ({f_i}(T_i^*) - {c_i}(T_i^*)) + (1 - O(\xi )){c_i}(T_i^*)] \\ 
  & = \sum\limits_{i = 1}^m {[\frac{1}{p}(1 - {e^{ - p}})} {f_i}(T_i^*) + (1 - \frac{1}{p} + \frac{{{e^{ - p}}}}{p} - O(\xi )){c_i}(T_i^*)] \notag\\ 
  & \ge \sum\limits_{i = 1}^m {[\frac{1}{p}(1 \! - \! {e^{ - p}})} {f_i}(T_i^*) \! + \! (1 \! - \! \frac{1}{p} \! + \! \frac{{{e^{ - p}}}}{p} \! - \! O(\xi ))(1 \! - \! {\kappa _{f1}} \! + \! O({c_2})){f_i}(T_i^*)] \notag\\ 
  & = [1 - \frac{{{\kappa _{f1}}{e^{ - p}}}}{p} + \frac{{{\kappa _{f1}}}}{p} - {\kappa _{f1}} + O(\xi ,{c_2})] \cdot \sum\limits_{i = 1}^m {{f_i}(T_i^*)}  \notag 
\end{align}
\end{scriptsize}

\vspace{-10pt}

\noindent where $O(\xi ,{c_2}) = ({\kappa _{f1}} - 1)O(\xi ) + [1 - \frac{1}{p} + \frac{{{e^{ - p}}}}{p} - O(\xi )]O({c_2})$ and ${S^*} = \bigcup\nolimits_i {T_i^*}$. After inequality (\ref{fun:ineq1}) is divided by $m$, we can obtain the final result. \hfill $\square$\par

Note that it is acceptable to claim qualitatively that the algorithms with higher approximation guarantee usually have a better objective value than those with lower guarantee \cite{stan2017probabilistic}\cite{krause2014submodular}. For approximation guarantee in inequality (\ref{ieq:TSSOguarantee}), it contains the pure guarantee form $1 - \frac{{{\kappa _{f1}}{e^{ - p}}}}{p} + \frac{{{\kappa _{f1}}}}{p} - {\kappa _{f1}}$ similar to the traditional one \cite{fujishige2005submodular}, and an error form $O(\xi ,{c_2}) \leq 0$. From their formulation, as the cardinality of $|S_{\rm{L}1}|$ increases, the separate curvature $\kappa _{f1}$ decreases, implying that the pure guarantee will increase. Unluckily, increasing $|S_{\rm{L}1}|$ may worsen $O(\xi ,{c_2})$ determined by both $\kappa _{f1}$ and $\kappa _{f2}$. By coordinating the pure guarantee and error, Algorithm 1 can generate a better solution than other TSSO solving methods. Furthermore, under some $\kappa _{f1}$ and $p$, the value of pure guarantee can be larger than traditional one $1 - {e^{ - 1}}$, and its guarantee ratio distribution is illustrated in Fig. \ref{fig:cur}.

Since there is only one constraint in second stage for TSSO in this research, we can obtain Corollary 1.

\noindent \textbf{Corollary 1.} With $p=1$, the approximation ratio from Theorem 3 can be simplified as

\vspace{-13pt}

\begin{small}
\begin{align} \label{ieq:TSSO}
    F({S^k}) \ge [1 - {\kappa _{f1}}{e^{ - 1}} + O'(\xi ,{c_2})] \cdot F({S^*})
\end{align}
\end{small}

\vspace{-8pt}

\noindent \textbf{Proof:} When $p=1$ is introduced into Theorem 3, it is easy to obtain the result, where $O'(\xi ,{c_2}) = ({\kappa _{f1}} - 1)O(\xi ) + [{e^{ - 1}} - O(\xi )]O({c_2})$.  \hfill $\square$\par

\noindent \textbf{Theorem 4.} Let $k$ denote the first stage constraint, ${k_m} = \max {k_{c2}}$ the maximum constraint in second stage, $D = {\max _i}|{M^i}|$ the maximum size of sub-database, ${d_m} = {\max _i}{\max _j}|{M^{i(j)}}|$ the generation maximum, $m$ the number of system state and $n = |{S_{\rm{L}}}|$ the line candidate number. Then the optimization of Algorithm 1 runs in $O(k{k_m}n^{2}mD{d_m})$ time.

\noindent \textbf{Proof:} The runtime of Algorithm 1 is decided by the searching processes in first and second stages of TSSO and the calculation of sub-function $f(S)$. During the first stage searching process, the algorithm iteratively scans $n$ components at most $k$ times, i.e., $O(nk)$. In the second stage, there are $m$ sub-functions with maximized constraint $k_m$, i.e., $O(m k_m)$. In addition, each sub-function involves at most $O(n D d_m)$ extraction of the failure data. Finally, the total complexity of Algorithm 1 is $O(k{k_m}n^{2}mD{d_m})$. \hfill $\square$\par

Theorem 4 implies that the dedicated solving algorithm can handle TSSO in polynomial time. In this study, the computation efficiency is compared to the original exponential computation complexity in tackling the identical problem. When dealing with large-scale cases, our method’s scalability is more evident compared to traversal process with at least $O(2^n)$ complexity, particularly in large-scale examples prone to the dimensionality curse \cite{krause2014submodular}.

\vspace{-10pt}

\section{Case Studies}
\subsection{Impacts of environment and separate curvature on DTR risk mitigation}
According to the regulation \cite{iglesias2014guide}, the value of DTR is dependent to ambient environment factors such as wind speed $V$ and surrounding temperature $Te$. Also, in order to determine the reliable DTR for relay threshold, it is suitable to use the worst environment information among all the critical spans of a transmission line \cite{xiao2018power}. Thus, the first objective in this part is to investigate the effect of environment factor, represented by the DTR improved factor $\alpha$, on DTR risk mitigation performance.

Experiments are operated in IEEE 39-bus system. Specifically, suppose that there are 10 system states, and the constraints in the first and second stages are $k_{c1} =8$ and $k_{c2}=[3,4,4,3,3,3,3,4,3,3]$, respectively. Set $\eta=0.5$  and $Y_{ext}=1000$ for submodular sub-function, and each sub-database $M^l$ contains 2000 simulations. Assume that the default threshold rating is from the condition that $Te=40$°C and $V=0.61m/s$ \cite{iglesias2014guide}, implying that $\alpha=1.0$. Then we select the average worst weather data in different critical line spans from weather databases MRCC \cite{cli-MATE1} and NOAA \cite{NOAAs19812010USClimateNormalsAnOverview} to calculate $\alpha$.

\begin{table}[!t] \scriptsize
\setlength{\abovecaptionskip}{0cm}
\renewcommand{\arraystretch}{1.3}
\captionsetup{font={footnotesize}}
\caption{WEATHER EFFECT ON DTR RISK MITIGATION}
\label{table:Weather}
\centering
\resizebox{\linewidth}{!}{
\begin{tabular}{ccccc}
\hline
Average worst weather  & $\alpha$  & $F$   & $RiskW$   & $BPI$    \\ \hline
$Te=40$, $V=0.61$         & 1    & 0.000    & 2822.300 & 0.000   \\
$Te =38.5$, $V =1.29$     & 1.03 & 506.181  & 1944.031 & 161.548 \\
$Te =37.2$, $V =1.76$     & 1.05 & 940.643  & 1543.531 & 229.475 \\
$Te =36.8$, $V =1.79$     & 1.07 & 1127.598 & 1339.127 & 194.576 \\
$Te =36.3$, $V =1.85$     & 1.09 & 1395.570 & 1083.234 & 218.735 \\
$Te =35.7$, $V =1.94$     & 1.11 & 1614.099 & 836.569  & 162.462 \\ \hline
\end{tabular}%
}
\vspace{-20pt}
\end{table}

Table \ref{table:Weather} demonstrates the simulation results. With a better weather condition, $\alpha$ value increases, indicating that more remaining transmission capacity can be exploited with the help of DTR. When $\alpha$ rises, the risk mitigation value $F$ becomes larger, raising from $F=506.181$ as $\alpha=1.03$ to $F=1127.598$ as $\alpha=1.07$, and $F=1614.099$ as $\alpha=1.11$. Similarly, the risk value $RiskW$ drops from the original risk $RiskW=2822.300$ at $\alpha=1$ to $RiskW=1543.531$ at $\alpha=1.05$, and further declines to $RiskW=836.569$ at $\alpha=1.11$. These indexes exhibit that the risk mitigation effect will become larger with a better weather condition, i.e., a larger $\alpha$ value. What’s more, there is Braess paradox happening in risk mitigation, as shown by $BPI \in (160,230)$ with DTR placement in system. It reminds us that it is better to apply the DTR to mitigate the failure risk meanwhile keeping the Braess paradox effect at an acceptable level.

On the other hand, the separate sets $S_{\rm{L}1}$ and $S_{\rm{L}2}$ can also impact the curvature value, hence affecting the approximation guarantee. Based on in inequalities (\ref{ieq:TSSO}), the roles of $\kappa _{f1}$ and $O'(\xi ,{c_2})$ are somehow competitive in guarantee improvement, so we should select $S_{\rm{L}1}$ carefully, considering both cardinality and its components, to achieve a better approximation guarantee. The settings are the same as above besides fixed $\alpha=1.05$, and Table \ref{table:curv} shows the influence of separate curvature on risk mitigation.

\begin{table}[!t] \scriptsize
\setlength{\abovecaptionskip}{0cm}
\renewcommand{\arraystretch}{1.3}
\captionsetup{font={footnotesize}}
\caption{SEPARATE CURVATURE EFFECT ON DTR RISK MITIGATION}
\label{table:curv}
\centering
\begin{tabular}{cccccc}
\hline
$|S_{\rm{L1}}|$ & $F$   & $\kappa _{f1}$   & $Pure Guarantee$   & $O'(\xi ,{c_2})$  & $Guarantee$  \\ \hline
11 & 750.148 & 0.510 & 0.812          & -0.314 & 0.498     \\
15 & 827.283 & 0.536 & 0.803          & -0.384 & 0.419     \\
18 & 771.792 & 0.665 & 0.755          & -0.213 & 0.542     \\
21 & 752.483 & 0.674 & 0.752          & -0.489 & 0.262     \\
24 & 772.430 & 0.687 & 0.747          & -0.588 & 0.159     \\
27 & 810.735 & 0.783 & 0.712          & -0.111 & 0.601     \\
30 & 846.226 & 0.697 & 0.744          & -0.199 & 0.545     \\
33 & 880.912 & 0.837 & 0.692          & -0.139 & 0.554     \\
36 & 940.643 & 0.765 & 0.719          & -0.045 & 0.674     \\
38 & 941.036 & 0.675 & 0.752          & -0.163 & 0.589     \\
41 & 901.453 & 0.735 & 0.730          & -0.166 & 0.564     \\
45 & 889.956 & 0.667 & 0.755          & -0.203 & 0.552     \\ \hline
\end{tabular}%
\vspace{-10pt}
\end{table}

\begin{table}[!t] \scriptsize
\setlength{\abovecaptionskip}{0cm}
\renewcommand{\arraystretch}{1.3}
\captionsetup{font={footnotesize}}
\caption{PERFORMANCE COMPARISON BETWEEN ONE-STAGE AND TWO-STAGE MODELS}
\label{table:Comp2One}
\centering
\begin{tabular}{cccccc}
\hline
\multirow{2}{*}{System
state} & \multicolumn{2}{c}{One-stage
model} & \multicolumn{3}{c}{Two-stage
model}     \\ 
\cline{2-6}
                              & \textit{f} & \textit{BPI}           & \textit{f} & \textit{BPI} & \textit{T}  \\ 
\hline
1                             & 151.809    & 71.268                 & 105.301    & 28.972       & 11,16,23    \\
2                             & 340.600    & 379.013                & 454.316    & 29.772       & 9,16,19,23  \\
3                             & 771.740    & 219.062                & 1347.633   & 106.127      & 3,9,11,16   \\
4                             & 792.133    & 28.022                 & 922.339    & 28.577       & 3,9,23      \\
5                             & 1088.631   & 0.000                  & 1378.213   & 0.000        & 9,16,19     \\
6                             & 868.428    & 524.382                & 981.752    & 410.267      & 9,19,27     \\
7                             & 1320.952   & 257.718                & 1265.288   & 264.595      & 3,9,27      \\
8                             & 258.908    & 561.247                & 341.233    & 504.690      & 3,11,16,23  \\
9                             & 1313.603   & 488.344                & 1448.525   & 396.493      & 11,19,27    \\
10                            & 855.424    & 290.773                & 1161.833   & 525.254      & 3,16,45     \\
Mean                          & 776.223    & 281.983                & 940.643    & 229.475      & -           \\
\hline
\end{tabular}%
\vspace{-20pt}
\end{table}

We can see that when separate number $|S_{\rm{L1}}|$ decreases, the pure guarantee grows, while there is also a worser $O'(\xi ,{c_2})$, resulting in a poor guarantee. For instance, even if $1 - {\kappa _{f1}}{e^{ - 1}}=0.803$ when $|S_{\rm{L}1}|=15$, the worser error $O'(\xi ,{c_2})=-0.384$ yields a lower guarantee value 0.419. As $|S_{\rm{L}1}|$ rises, both pure guarantee and $O'(\xi ,{c_2})$ have downward trends, but $O'(\xi ,{c_2})$ decrease more. Then, we can find a key balance between pure guarantee and error form to obtain a larger guarantee. In this experiment, when $|S_{\rm{L}1}|=36$, the guarantee is 0.674, which is more than the traditional optimal guarantee $1-e^{-1}$ \cite{fujishige2005submodular}, and its objective value $F=940.643$ is nearly the best compared to other $|S_{\rm{L}1}|$. Note that with $|S_{\rm{L}1}|=38$, the objective value $F=941.036$ is close to the value at $|S_{\rm{L}1}|=36$, but its guarantee is 0.589 lower than guarantee at $|S_{\rm{L}1}|=36$, suggesting qualitatively that under some extreme conditions, like bad or missing state data, the worst value at $|S_{\rm{L}1}|=38$ will be less than the worst value at $|S_{\rm{L}1}|=36$. 

In summary, the DTR can indeed assist to mitigate the risk. Regarding the weather factor, DTR can release more remaining transmission capacity with a better average weather condition. The separate curvature, on the other hand, can influence the DTR placement and operation scheme as well. Careful selection of separate sets can reach a suitable balance between pure guarantee and error form, leading to a superior performance guarantee.

\vspace{-10pt}

\subsection{Performance comparison with one-stage optimization}
In this part, the goal is to compare the performances between one-stage and two-stage DTR models. The comparison experiment is conducted in IEEE 39-bus system. The settings of one-stage and two-stage model are the same as above and fix $\alpha=1.05$. But the difference is that one-stage model only contains constraint $k=5$, meaning that all the placed DTR in one-stage model will operate constantly regardless of the system state.

\begin{table*}[!t] \scriptsize
\setlength{\abovecaptionskip}{0cm}
\renewcommand{\arraystretch}{1.3}
\captionsetup{font={footnotesize}}
\caption{RESIDUAL SERVICE LIFE RATIO BETWEEN ONE-STAGE AND TWO-STAGE MODELS}
\label{table:Servicelife}
\centering
\begin{tabular}{c c c c c c c c c ccccc}
\hline
\multirow{2}{*}{Operation
year} & \multicolumn{5}{c}{One-stage
  model} & \multicolumn{8}{c}{Two-stage model~ ~ ~}               \\
  \cline{2-14}
                                & 3    & 6    & 9    & 16   & 27        & 3    & 9    & 11   & 16   & 19   & 23   & 27   & 45    \\ 
\hline
2                               & 0.67 & 0.67 & 0.67 & 0.67 & 0.67      & 0.83 & 0.80 & 0.87 & 0.80 & 0.87 & 0.87 & 0.90 & 0.97  \\
4                               & 0.33 & 0.33 & 0.33 & 0.33 & 0.33      & 0.67 & 0.60 & 0.73 & 0.60 & 0.73 & 0.73 & 0.80 & 0.93  \\
\hline
\end{tabular}%
\vspace{-17pt}
\end{table*}

\begin{table}[!t] \scriptsize
\setlength{\abovecaptionskip}{0cm}
\renewcommand{\arraystretch}{1.3}
\captionsetup{font={footnotesize}}
\caption{PERFORMANCE COMPARISON UNDER LOAD GROWTH SITUATION}
\label{table:ComLoad}
\centering
\begin{tabular}{ccccc}
\hline
\multirow{2}{*}{LoadR} & \multicolumn{2}{c}{1.02}  & \multicolumn{2}{c}{1.06}                    \\ 
\cline{2-5}
                       & \textit{F}  & \textit{BPI} & \textit{F}  & \textit{BPI}  \\ 
\hline
One-stage              & 805.863       & 257.038      & 852.072      & 220.321       \\
Two-stage              & 934.957       & 250.579      & 950.992      & 240.795       \\
Flexible               & 982.375       & 247.480      & 991.318      & 237.968       \\
\hline
\end{tabular}%
\vspace{-10pt}
\end{table}

The result is shown in TABLE \ref{table:Comp2One}. In one-stage model, the selected lines are 3, 6, 9, 16, 27, and that in two-stage model are 3, 9, 11, 16, 19, 23, 27, 45. As can be seen, two-stage model enables DTR operating flexibly with redundant placement, which can reach a better effect. For instance, even though two-stage model has fewer DTR operation in state 3, $f^{two}=1645.468$ is greater than $f^{one}=771.740$ in one-stage model. The similar results also occur in states 2, 4, 5, 6, 8, 9, 10. And from $BPI$ value, the fixed DTR operation in one-stage model can cause more Braess paradox. In state 2, for example, $BPI^{one}=379.013$ is larger than $BPI^{two}=29.772$, meaning that one-stage model inversely brings more extra risk into system. For the whole performance, mean value $F^{one}=776.223$ is less than $F^{two}=940.643$, while mean value $BPI^{one}=281.983$ is worser than $BPI^{two}=229.475$. These results demonstrate that two-stage model can schedule fewer DTR properly to achieve a higher risk mitigation effect. Since there is only one stage optimization, one-stage model needs to balance the risk mitigation effect in each system state, so that the final DTR operation is not the best scheme for all states.

Moreover, we contrast the residual service life of DTR in 2 models. Assume that each DTR can work around the clock for 6 years, then the residual service life ratio for each DTR can be calculated, shown in TABLE \ref{table:Servicelife}. From the result, the residual service life ratio for each DTR in one-stage model is only 67\% after 2 years and 33\% after 4 years. However, after 2 years and 4 years, the residual service life ratios for each line in two-stage model are all larger than 80\% and 60\%, respectively, indicating that the flexible DTR operation can extend the DTR service life.

Furthermore, we also investigate the two-stage model’s flexibility to potential load increase. In addition to the one-stage and two-stage model defined before, we include a flexible two-stage model that adds an extra DTR in each system state. Setting load ratios $LoadR$ are 1.02 and 1.06 times of the original, respectively. The result in TABLE \ref{table:ComLoad} shows that, as system load grows the two-stage model performs better in risk mitigation than one-stage model, implied by that $F^{two}=934.957$ is larger than $F^{one}=805.863$ at $LoadR=1.02$, and by the similar results at $LoadR=1.06$. In comparison to one-stage and original two-stage models, flexibility of redundant DTR allows the flexible two-stage model to achieve the highest risk mitigation effect, demonstrated by $F^{flex}=982.375$ at $LoadR=1.02$ and $F^{flex}=991.318$ at $LoadR=1.06$. It indicates that the two-stage model have scalability potential to handle the load growth and other emergencies, achieving a more flexible and effective performance of risk mitigation.

To sum up, two-stage method considers the suitable redundant DTR placement, not only allowing operators to set up the flexible DTR operation schedule for a specific system state, but also prolonging DTR service life. It can also function better when faced with load growth and other situations, showing advantages compared to one-stage model.

\vspace{-10pt}

\subsection{Performance comparison with different strategies}
This section compares the performance of our proposed strategy, \textbf{separate curvature greedy strategy (SCG)} with that of other two-stage strategies. Through the literature review, we take 3 categories as the comparable forms.

The first category is known as the \textbf{index-based strategy}, which selects DTR lines based upon some traditional assessment indexes \cite{long2021analyzing}\cite{xu2020mitigating}, including:

\begin{itemize} \small

\item[$\bullet$] \textbf{Random line strategy (RL)}. Select lines randomly to place DTR.
\item[$\bullet$] \textbf{Failure rate strategy (FR)}. Sort lines in decreasing order from failure number in database, and place DTR in the highest ranked lines.
\item[$\bullet$] \textbf{Largest power flow strategy (LPF)}. Sort lines in decreasing order from initial power flow, and place DTR in the highest ranked lines.
\item[$\bullet$] \textbf{Largest hidden failure strategy (LHF)}. Using $N-1$ security test, sort lines in decreasing order from hidden failure probabilities, and place DTR in the highest ranked lines.

\end{itemize}

The second category, known as \textbf{greed-based strategy}, includes the core greedy algorithm but does so in a variant way \cite{balkanski2016learning}\cite{sviridenko2017optimal}. It includes:

\begin{itemize} \small

\item[$\bullet$] \textbf{Greedy sum strategy (GS)}. Select $k$ lines with the highest objective function $F$ iteratively, then choose the matching constrained lines for each system state from these $k$ lines.
\item[$\bullet$] \textbf{Modular approximation strategy (MA)}. Approximate the sub-function $f$ is modular, then select $k$ lines with the highest objective function $F$ iteratively.
\item[$\bullet$] \textbf{Local search strategy (LS)}. Select a line with the highest objective function $F$ and arbitrary $k-1$ lines, then choose an unselected line to replace the selective line constantly. If the new objective value is larger than the previous, adopt this replacement set. Note that it can achieve a $\frac{1}{{p + 1}}$  guarantee where $p$ is second stage constraint’s number.

\end{itemize}

The last category is called \textbf{surrogate-based strategy}, which designs several surrogate functions to judge DTR line selection, including: 

\begin{itemize} \small

\item[$\bullet$] \textbf{Replacement greedy strategy (RG)}. Select $k$ lines with the highest surrogate function $f({S_{\rm{A}}} \cup s) - f({S_{\rm{A}}})$ iteratively \cite{stan2017probabilistic}. And it can achieve a $\frac{{1 - {e^{ - (p + 1)}}}}{{p + 1}}$ guarantee.
\item[$\bullet$] \textbf{General P-matroid greedy strategy (GPG)}. Select $k$ lines with the highest surrogate function ${(1 - \frac{2}{k})^{k - i}}g({S_{\rm{A}}}) + c({S_{\rm{A}}})$ iteratively, where $k$ and $i$ denote the first stage constraint and searching step \cite{yang2021constrained}. And it can achieve the same guarantee as RG.
\item[$\bullet$] \textbf{General curvature-based greedy strategy (GCG)}. Select $k$ lines with the highest surrogate function ${(1 - \frac{{p + 1}}{k})^{k - i}}g({S_{\rm{A}}}) + {(1 - \frac{p}{k})^{k - i}}c({S_{\rm{A}}})$ iteratively \cite{li2021two}. And it can achieve a $\frac{{1 - {\kappa _f}}}{p}(1 - {e^{ - p}}) + \frac{{{\kappa _f}}}{{p + 1}}(1 - {e^{ - (p + 1)}})$ guarantee where $\kappa _f$ is curvature.

\end{itemize}

The experiment settings are the same as in $Section$ $VI.B$ and the result is given in TABLE \ref{table:Comp2Diff}. For index-based strategies, they perform worse than other categories even though LHF has a higher objective value $F^{\rm{LHF}}=743.385$, since the indexes in these strategies are not closely correlated to the risk and capture inadequate features in the cascading failure process. For greed-based strategies, they all have relatively better performance, shown in $F^{\rm{GS}}=872.411$, $F^{\rm{MA}}=856.520$ and $F^{\rm{LS}}=886.701$. And for surrogate-based strategy, RG performs better than greed-based strategies as $F^{\rm{RG}}=890.748$, while other 2 perform poorly. The reason is that, while our analyzed problem only contains a single cardinality constraint, GPG and GCG is designed for the general situation and perform better in complex constraints \cite{li2021two}\cite{yang2021constrained}, as opposed to the index-based strategy and greed-based strategy which are only intended to be used for single constraint type. For our strategy SCG, it has the best performance $F^{\rm{SCG}}=940.643$ with adaptive regulation of separate curvature to theoretically boost the approximation guarantee compared to others. What’s more, the $BPI$ value does not equal zero in all strategies, serving as a reminder that the Braess paradox must be considered when designing a risk mitigation scheme.

\begin{table}[!t] \scriptsize
\setlength{\abovecaptionskip}{0cm}
\renewcommand{\arraystretch}{1.2}
\captionsetup{font={footnotesize}}
\caption{PERFORMANCE COMPARISON AMONG DIFFERENT STRATEGIES}
\label{table:Comp2Diff}
\centering
\begin{tabular}{ccc}
\hline
Strategy & $F$   & $BPI$    \\ \hline
RL         & 488.998 & 140.033                           \\
FR         & 595.917 & 271.307                          \\
LPF        & 31.496  & 3.942                          \\
LHF        & 743.385 & 243.807                        \\ 
\cdashline{1-3}[0.2pt/5pt]
GS         & 872.411 & 293.569                          \\
MA         & 856.520 & 263.058                          \\
LS         & 886.701 & 189.308                          \\ \cdashline{1-3}[0.2pt/5pt]
RG         & 890.748 & 201.953                          \\
GPG        & 497.852 & 136.508                          \\
GCG        & 465.902 & 175.473                          \\ \cdashline{1-3}[0.2pt/5pt]
SCG (ours) & 940.643 & 229.475                          \\
No DTR     & 0.000   & 0.000                           \\ \hline
\end{tabular}%
\vspace{-20pt}
\end{table}

\begin{figure}[!t]
    \centering
    \includegraphics[width=8cm]{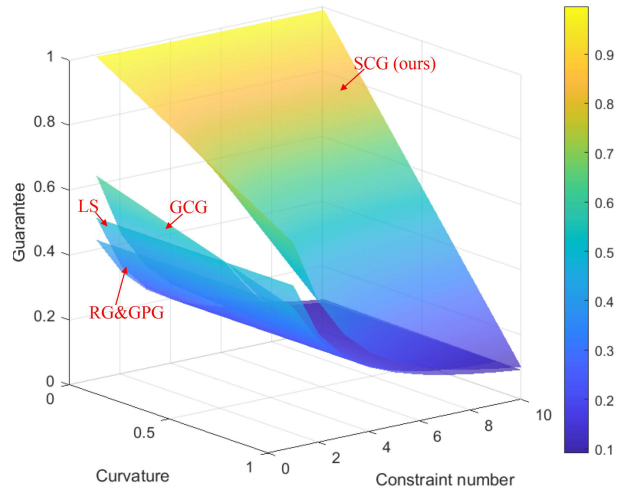}
    \captionsetup{font={footnotesize}}
    \caption{Guarantee comparison in TSSO among different Strategies}
    \label{fig:cur}
    \vspace{-20pt}
\end{figure}

Moreover, Fig. \ref{fig:cur} compares the guarantee distributions among LS, RG, GPG, GCG and SCG, neglecting their error forms. When curvature or constraint number lowers, we can see that the guarantees of all strategies improve. With the help of separate curvature, the proposed strategy SCG can reach a higher performance under the same conditions compared to others. And it is striking that SCG’s guarantee may be raised close to 1, showing its greater improvement potential.

In conclusion, the proposed strategy SCG outperforms other strategies in TSSO-based risk mitigation. With the aid of separate curvature, SCG can improve the objective function value adaptively, seen from the highest approximation guarantee. And due to the consideration of constraint’s number, SCG is also suitable to multiple constraint types.

\section{Conclusion}
This work investigated a DTR optimization problem for risk mitigation considering placement and operation schedule in terms of service life and Braess paradox. The proposed model is based on two-stage submodular optimization, in which the DTR placement with proper redundance is optimized in first stage and the flexible operation schedule is designed for each system state in second stage. First, we established a sub-function of CF risk mitigation. Then using the Markov and submodular properties in sub-function, the submodularity of total objective function of TSSO is proven for the first time. Consequently, a state-of-the-art solving algorithm based on separate curvature is devised, which can provide a better provable approximation guarantee than the current researches and obtain the solution in polynomial time. Case results demonstrate the impacts of environment and separate curvature on risk mitigation. Also, it validates that the suggested model outperforms the one-stage model both in risk mitigation and prolonging service life, and exceeds other two-stage strategies in performance.


%

\appendix
\noindent \textit{\textbf{Proof of the Submodularity of TSSO with Markov property}}
To simplify the notation, we apply set function $Y_i (T_i )=Cons_i-y_i (T_i)$ to represent the sub-function $f_i (T_i)$, where $Cons_i$ is a constant, $y_i (T_i )$ is a Markov-based decreasing function having $y_i (a \cup b)=y_i (a) \cdot y_i (b)$ where $a \cap b=\emptyset$, and $i$ is the sub-function index. Note that $Y_i (T_i ) \geq 0$, and $y_i (T_i )$ can be treated as a general form of risk mitigation $RiskW_{T_i}^i$ (the smaller the better). Define $G_i (S)=\{T \subseteq S:T=argmax f_i (\cdot)\}$ as the optimal solution for $y_i (T_i)$ in the feasible region $S$, and $S_{\rm{L}}$ the ground set. Then TSSO can be formulated by the weight sum of $Y_i (G_i (S))$, which chooses the feasible region $S$ for sub-functions $Y_i (T_i)$ to get the optimal results. Therefore, the goal of checking the submodularity of TSSO under Markov property is to show whether the $Y_i (G_i (S))$ is submodular. For simplicity, we can analyze the submodularity of $y_i (G_i (S))$, which allows us to easily infer the $Y_i (G_i (S))$ properties.

In the beginning, Lemma A1 is introduced to clarify the inequality properties of $y_i (G_i (S))$.

\noindent \textbf{Lemma A1.} For Markov-based function $y_i (T_i )$, if there are subsets $S_{\rm{A}}$, $S_{\rm{B}}$, $v$ satisfying $S_{\rm{A}}  \subseteq S_{\rm{B}} \subseteq S_{\rm{L}}$, $v\in S_{\rm{L}} \backslash S_{\rm{B}}$, $y_i (T_i )$ has these inequalities in combinatorial optimization: ${y_i}({G_i}({S_{\rm{A}}})) \ge {y_i}({G_i}({S_{\rm{A}}} \cup v))$, ${y_i}({G_i}({S_{\rm{B}}})) \ge {y_i}({G_i}({S_{\rm{B}}} \cup v))$, ${y_i}({G_i}({S_{\rm{A}}})) \ge {y_i}({G_i}({S_{\rm{B}}}))$ and ${y_i}({G_i}({S_{\rm{A}}} \cup v)) \ge {y_i}({G_i}({S_{\rm{B}}} \cup v))$. Note that for $y_i (T_i )$, the smaller the better, and $Y_i (T_i )$ has inversive inequalities from $y_i (T_i )$.

\noindent \textbf{Proof:} For $y_i (G_i (S_{\rm{A}} )) \geq y_i (G_i (S_{\rm{A}} \cup v))$, based on the definition of $G_i (S)$, the adding of $v$ to $S_{\rm{A}}$ constructs a new feasible region, including 2 scenarios.

1) $G_i (S_{\rm{A}} \cup v)$ cannot diminish the value $y_i (G_i (S))$ by adding $v$ to $G_i (S_{\rm{A}} )$ or replacing certain components in $G_i (S_{\rm{A}} )$ with $v$, then it has $G_i (S_{\rm{A}} \cup v)=G_i (S_{\rm{A}} )$ such that $y_i (G_i (S_{\rm{A}} \cup v))=y_{i} (G_{i} (S_{\rm{A}} ))$.

2) $G_i (S_{\rm{A}} \cup v)$ reduces the value $y_i (G_i (S))$, meaning that $y_i (v) < y_i (G_i (S_{\rm{A}} ) \backslash G_i (S_{\rm{A}} \cup v))$, so that it has $y_i (G_i (S_{\rm{A}} \cup v)) < y_{i} (G_{i} (S_{\rm{A}} ))$.

Thus, there is $y_{i} (G_{i} (S_{\rm{A}} )) \geq y_i (G_i (S_{\rm{A}} \cup v))$. Other inequalities can be obtained by the similar process as described above. 

\hfill $\square$\par

Then some operations are defined for $G_i (S)$. 

When $G_i (S_{\rm{A}} \cup v) = G_i (S_{\rm{A}} )$ and there is a set $S_{\rm{B}}$ where $S_{\rm{B}} \neq S_{\rm{A}}$, we define Absorb operation (Abs-ope): ${G_i}({S_{\rm{B}}} \cup {S_{\rm{A}}} \cup v) = {G_i}({S_{\rm{B}}} \cup {S_{\rm{A}}} )$.

When $G_i (S_{\rm{A}} \cup v) \neq G_i (S_{\rm{A}} )$, we define three operations in set $G_i (S_{\rm{A}} \cup v)$:

\textcircled{1} Addition operation (Add-ope): $G_i (S_{\rm{A}} \cup v) = G_i (S_{\rm{A}} ) \cup v$.

\textcircled{2} Exchange operation (Exc-ope): $G_i (S_{\rm{A}} \cup v) = G_i (S_{\rm{A}} \backslash a ) \cup v$ where $a$ is the element in $G_i (S_{\rm{A}} )$ replaced by $v$ and $|a|=1$.

\textcircled{3} Reduction operation (Red-ope): $G_i (S_{\rm{A}} \cup v) = G_i (S_{\rm{A}} \backslash a ) \cup v$ where $a$ are the elements in $G_i (S_{\rm{A}} )$ replaced by $v$ and $|a|>1$.

After that, there are some important set relationships of $G_i (S)$ shown in Lemmas A2, A3 and A4.

\noindent \textbf{Lemma A2.} For subsets $S_{\rm{A}}$, $S_{\rm{B}}$, $v$ satisfying $S_{\rm{A}}  \subseteq S_{\rm{B}} \subseteq S_{\rm{L}}$, $v\in S_{\rm{L}} \backslash S_{\rm{B}}$,
$G_i (S_{\rm{A}} \cup v)=G_i (S_{\rm{A}} )$ leads to $G_i (S_{\rm{B}} \cup v)=G_i (S_{\rm{B}} )$.

\noindent \textbf{Proof:} Since $G_i (S_{\rm{A}} \cup v)=G_i (S_{\rm{A}} )$, it signifies that the adding of component $v$ to current feasible region $S_{\rm{A}}$ cannot decrease $y_i (G_i (S))$. Due to $y_{i} (G_{i} (S_{\rm{A}} )) \geq y_{i} (G_{i} (S_{\rm{B}} ))$, we know that the improvement $y_{i} (G_{i} (S_{\rm{B}} )) - y_{i} (G_{i} (S_{\rm{A}} ))$ is from components in set $S_{\rm{B}} \backslash S_{\rm{A}}$, not in $v$. Based upon Abs-ope, there is

\begin{align}
    &\quad {G_i}({S_{\rm{B}}} \cup v) \notag\\ 
    & = {G_i}({S_{\rm{B}}} \cup v \cup {S_{\rm{A}}})  \notag\\ 
    & = {G_i}({S_{\rm{B}}} \cup ({S_{\rm{A}}} \cup v))  \notag\\ 
    &= {G_i}({S_{\rm{B}}} \cup {S_{\rm{A}}}) \notag\\ 
    & = {G_i}({S_{\rm{B}}})
\end{align}

\hfill $\square$\par

\noindent \textbf{Lemma A3.} For subsets $S_{\rm{A}}$, $S_{\rm{B}}$, $v$ satisfying $S_{\rm{A}}  \subseteq S_{\rm{B}} \subseteq S_{\rm{L}}$, $v\in S_{\rm{L}} \backslash S_{\rm{B}}$, $G_i (S_{\rm{B}} \cup v) \neq G_i (S_{\rm{B}} )$ leads to $G_i (S_{\rm{A}} \cup v) \neq G_i (S_{\rm{A}} )$.

\noindent \textbf{Proof:} Since $G_i (S_{\rm{B}} \cup v) \neq G_i (S_{\rm{B}} )$, we can obtain that $y_i (G_i (S_{\rm{B}} \cup v)) < y_i (G_i (S_{\rm{B}} ))$, showing that the reduced value is contributed from component $v$. Due to $S_{\rm{A}}  \subseteq S_{\rm{B}}$, the adding of $v$ can also reduce the value of $y_{i} (G_{i} (S_{\rm{A}} ))$, i.e., $y_{i} (G_{i} (S_{\rm{A}} )) > y_i (G_i (S_{\rm{A}} \cup v))$. Then, we can obtain $G_i (S_{\rm{A}} \cup v) \neq G_i (S_{\rm{A}} )$. \hfill $\square$\par

\noindent \textbf{Lemma A4.} For subsets $S_{\rm{A}}$, $S_{\rm{B}}$, $v$ satisfying $S_{\rm{A}}  \subseteq S_{\rm{B}} \subseteq S_{\rm{L}}$, $v\in S_{\rm{L}} \backslash S_{\rm{B}}$, it exists that

1) If $G_i (S_{\rm{B}} \cup v)$ executes Add-ope from $G_i (S_{\rm{B}} )$, $G_i (S_{\rm{A}} \cup v)$ will perform Add-ope from $G_i (S_{\rm{A}} )$.

2) If $G_i (S_{\rm{B}} \cup v)$ executes Exc-ope from $G_i (S_{\rm{B}} )$, $G_i (S_{\rm{A}} \cup v)$ will perform Add-ope or Exc-ope from $G_i (S_{\rm{A}} )$.

3) If $G_i (S_{\rm{B}} \cup v)$ executes Red-ope from $G_i (S_{\rm{B}} )$, $G_i (S_{\rm{A}} \cup v)$ will perform Add-ope, Exc-ope or Red-ope from $G_i (S_{\rm{A}} )$.

\noindent \textbf{Proof:} These statements are proven in the perspectives of cost performance and residual budget. From the definition, $G_i (S)$ selects components from $S$ that are close to the budget threshold or have greater cost performance, to optimize the objective function. Specifically, the cost nearing to the budget threshold indicates that there is still enough residual budget to apply more components, and the limitation is the insufficient feasible region. On the other hand, in the situation of greater cost performance, the improvement limit is component efficiency, and the budget does not allow for the addition of another component.

The above analysis suggests that execution of Add-ope means that there is still enough residual budget to add $v$ to $G_i (S)$ to optimize the objective value. And the executions of Exc-ope and Red-ope demonstrate that the residual budget is insufficient to add more component to the selected set, and the cost performance of $v$ is larger than some components in $G_i (S)$. 

Based upon Lemmas A1 and A3, $G_i (S_{\rm{B}} )$ has either less residual budget or higher cost efficiency compared with $G_i (S_{\rm{A}} )$. Thus, when $G_i (S_{\rm{B}} \cup v)$ performs Add-ope from $G_i (S_{\rm{B}} )$, $G_i (S_{\rm{A}} \cup v)$ must has a larger budget to execute Add-ope in statement 1). Inversely, if $G_i (S_{\rm{A}} \cup v)$ takes Exc-ope and Red-ope in this situation, it means that $G_i (S_{\rm{A}} )$ has a lower budget than $G_i (S_{\rm{B}} )$, which violates the formulation of $G_i (S)$.

Similarly, when $G_i (S_{\rm{B}} \cup v)$ executes Exc-ope from $G_i (S_{\rm{B}} )$, i.e., $G_i (S_{\rm{B}} \cup v)$ has used the cost of a single component in $G_i (S_{\rm{B}} )$ to replace with $v$, $G_i (S_{\rm{A}} \cup v)$ with larger budget will not replace more components in $G_i (S_{\rm{A}} )$ with $v$, such that $G_i (S_{\rm{A}} \cup v)$ will execute Add-ope or Exc-ope in statement 2). And the situation of the statement 3) is similar to the analysis of statement 2). \hfill $\square$\par

After obtaining Lemmas A1$ \sim $ A4, the submodularity of TSSO in Markov setting can be shown.

\noindent \textbf{Theorem A1.} Under the Markov and submodular features of sub-function, the total objective function of TSSO is submodular.

\noindent \textbf{Proof:} Suppose there are subsets $S_{\rm{A}}$, $S_{\rm{B}}$, $v$ satisfying $S_{\rm{A}}  \subseteq S_{\rm{B}} \subseteq S_{\rm{L}}$, $v\in S_{\rm{L}} \backslash S_{\rm{B}}$. $y_i (T_i )$ is a Markov-based decreasing function having $y_i (a \cup b)=y_i (a) \cdot y_i (b)$, and $y_i (G_i (S))$ can be represented as the optimal value of $y_i (T_i )$ under feasible region $S$. Due to the relationship of $S_{\rm{A}}$, $S_{\rm{B}}$ and $v$, there are $y_{i} (G_{i} (S_{\rm{A}} )) \geq y_i (G_{i} (S_{\rm{B}} ))$ and $y_{i} (G_{i} (S )) \geq y_i (G_i (S \cup v))$. 
Also define the general sampling weight as $\overset{\frown}{W}$ \cite{guo2017toward}, e.g., $\overset{\frown}{W}_{\rm{B}-\rm{A}}=  y_i (G_i (S_{\rm{B}} ) / y_i (G_i (S_{\rm{A}} )$. And it is easy to deduce that $\overset{\frown}{W}_{\rm{B}-\rm{A}} \leq 1$, $\overset{\frown}{W}_{\rm{A} \cup v - \rm{A} } \leq 1$  and so on. Note that the Markov property of sub-function allows for the application of sampling weight technique on $y_i (G_i (S))$ \cite{guo2017toward}.

Then, there are 3 situations for the searching of the optimal value in TSSO.

1) When $v$ is added to the current feasible region, it occurs that $G_i (S_{\rm{A}} \cup v)=G_i (S_{\rm{A}} )$ and $G_i (S_{\rm{B}} \cup v)=G_i (S_{\rm{B}} )$, as stated in Lemma A2. This suggests that the addition of $v$ cannot further optimize the objection value in the current feasible regions, both in $S_{\rm{A}}$ and $S_{\rm{B}}$. Based on Lemma A1, it can be shown that $\overset{\frown}{W}_{\rm{B} \cup v - \rm{A} \cup v} = \overset{\frown}{W}_{\rm{B}-\rm{A}} \leq 1$. Then there is

\begin{align}
 &\quad {y_i}({G_i}({S_{\rm{A}}})) - {y_i}({G_i}({S_{\rm{A}}} \cup v))  \notag\\ 
  &\ge {{\mathord{\buildrel{\lower3pt\hbox{$\scriptscriptstyle\frown$}} 
\over W} }_{{\rm{B - A}}}} \cdot {y_i}({G_i}({S_{\rm{A}}})) - {{\mathord{\buildrel{\lower3pt\hbox{$\scriptscriptstyle\frown$}} 
\over W} }_{{\rm{B}} \cup v - {\rm{A}} \cup v}} \cdot {y_i}({G_i}({S_{\rm{A}}} \cup v)) \notag\\ 
 & = {y_i}({G_i}({S_{\rm{B}}})) - {y_i}({G_i}({S_{\rm{B}}} \cup v))  
\end{align}

2) When $v$ is added to the current feasible region, it occurs that $G_i (S_{\rm{A}} \cup v) \neq G_i (S_{\rm{A}} )$ and $G_i (S_{\rm{B}} \cup v)=G_i (S_{\rm{B}} )$. It implies that $G_i (S_{\rm{A}} )$ has a larger budget to obtain $v$ than $G_i (S_{\rm{B}} )$, or $v$ has a higher cost performance for $G_i (S_{\rm{A}} )$ than $G_i (S_{\rm{B}} )$. In this situation,  $y_i (G_i (S_{\rm{A}} \cup v))$  will be less than $y_{i} (G_{i} (S_{\rm{A}} ))$ such that $\overset{\frown}{W}_{\rm{B} \cup v - \rm{A} \cup v} > \overset{\frown}{W}_{\rm{B}-\rm{A}} $. Then there is

\begin{align}
  &\quad {y_i}({G_i}({S_{\rm{A}}})) - {y_i}({G_i}({S_{\rm{A}}} \cup v)) \notag\\ 
  & \ge {{\mathord{\buildrel{\lower3pt\hbox{$\scriptscriptstyle\frown$}} 
\over W} }_{{\rm{B - A}}}} \cdot {y_i}({G_i}({S_{\rm{A}}})) - {{\mathord{\buildrel{\lower3pt\hbox{$\scriptscriptstyle\frown$}} 
\over W} }_{{\rm{B - A}}}} \cdot {y_i}({G_i}({S_{\rm{A}}} \cup v)) \notag\\ 
  &> {{\mathord{\buildrel{\lower3pt\hbox{$\scriptscriptstyle\frown$}} 
\over W} }_{{\rm{B - A}}}} \cdot {y_i}({G_i}({S_{\rm{A}}})) - {{\mathord{\buildrel{\lower3pt\hbox{$\scriptscriptstyle\frown$}} 
\over W} }_{{\rm{B}} \cup v - {\rm{A}} \cup v}} \cdot {y_i}({G_i}({S_{\rm{A}}} \cup v)) \notag\\ 
  &= {y_i}({G_i}({S_{\rm{B}}})) - {y_i}({G_i}({S_{\rm{B}}} \cup v)) 
\end{align}

3) When $v$ is added to the current feasible region, it occurs that $G_i (S_{\rm{A}} \cup v) \neq G_i (S_{\rm{A}} )$ and $G_i (S_{\rm{B}} \cup v) \neq G_i (S_{\rm{B}} )$, as stated in Lemma A3. This situation is complicated, which can be categorized into 3 cases indicated in Lemma A4.

Case (1). When $G_i (S_{\rm{B}} \cup v)$ executes Add-ope from $G_i (S_{\rm{B}} )$, $G_i (S_{\rm{A}} \cup v)$ will performs Add-ope from $G_i (S_{\rm{A}} )$. Due to the Markov property of $y_i (T_i )$, we can deduce that $y_i (G_i (S_{\rm{A}} \cup v)) = y_{i} (G_{i} (S_{\rm{A}} )) \cdot y_i (v)$ and $y_i (G_i (S_{\rm{B}} \cup v)) = y_{i} (G_{i} (S_{\rm{B}} )) \cdot y_i (v)$. Then there is $\frac{y_i (G_i (S_{\rm{A}} \cup v))}{y_{i} (G_{i} (S_{\rm{A}} ))} = \frac{y_i (G_i (S_{\rm{B}} \cup v))}{y_{i} (G_{i} (S_{\rm{B}} ))}$ such that $\overset{\frown}{W}_{\rm{B} \cup v - \rm{A} \cup v} = \overset{\frown}{W}_{\rm{B}-\rm{A}}$.

Case (2). When $G_i (S_{\rm{B}} \cup v)$ executes Exc-ope from $G_i (S_{\rm{B}} )$, $G_i (S_{\rm{A}} \cup v)$ will perform Add-ope or Exc-ope from $G_i (S_{\rm{A}} )$. In this case, $y_i (G_i (S_{\rm{B}} \cup v))= \frac{y_{i} (G_{i} (S_{\rm{B}} )) \cdot y_i (v)}{y_i (b)} $ where component $b$ in $G_i (S_{\rm{B}} )$ is replaced by $v$, implying that $y_i (v)<y_i (b)$. On the other hand, there is $y_i (G_i (S_{\rm{A}} \cup v)) = y_{i} (G_{i} (S_{\rm{A}} )) \cdot y_i (v)$ for Add-ope, or $y_i (G_i (S_{\rm{A}} \cup v))= \frac{y_{i} (G_{i} (S_{\rm{A}} )) \cdot y_i (v)}{y_i (a)} $ for Exc-ope, where component $a$ in $G_i (S_{\rm{A}} )$ is replaced by $v$. Due to the submodularity, if $G_i (S_{\rm{A}} \cup v)$ executes Add-ope, the reduction effect of $v$ on $y_{i} (G_{i} (S_{\rm{A}} ))$ must be greater than $y_{i} (G_{i} (S_{\rm{B}} ))$, leading to

\begin{align}
 &\quad {\rm{    }}\frac{{{y_i}({G_i}({S_{\rm{B}}})) \cdot {y_i}(v)}}{{{y_i}({G_i}({S_{\rm{B}}})) \cdot {y_i}(b)}} > \frac{{{y_i}({G_i}({S_{\rm{A}}})) \cdot {y_i}(v)}}{{{y_i}({G_i}({S_{\rm{A}}}))}} \notag\\ 
  & \Rightarrow \frac{{{y_i}({G_i}({S_{\rm{B}}} \cup v))}}{{{y_i}({G_i}({S_{\rm{B}}}))}} > \frac{{{y_i}({G_i}({S_{\rm{A}}} \cup v))}}{{{y_i}({G_i}({S_{\rm{A}}}))}} \notag\\ 
  & \Rightarrow {{\mathord{\buildrel{\lower3pt\hbox{$\scriptscriptstyle\frown$}} 
\over W} }_{{\rm{B}} \cup v - {\rm{A}} \cup v}} > {{\mathord{\buildrel{\lower3pt\hbox{$\scriptscriptstyle\frown$}} 
\over W} }_{{\rm{B - A}}}} 
\end{align}

If $G_i (S_{\rm{A}} \cup v)$ executes Exc-ope, the reduction effect from component $a$ is equal to or less than that from $b$, i.e., $y_i (b) \leq y_i (a)$. The reason is that the worse cost-effective component in $G_i (S_{\rm{B}} )$ can have equal or greater effect on reduction compared to the worse cost-effective component in $G_i (S_{\rm{A}} )$ due to $S_{\rm{A}}  \subseteq S_{\rm{B}}$. Then it has

\begin{align}
 &\quad \frac{{{y_i}({G_i}({S_{\rm{B}}})) \cdot {y_i}(v)}}{{{y_i}({G_i}({S_{\rm{B}}})) \cdot {y_i}(b)}} \ge \frac{{{y_i}({G_i}({S_{\rm{A}}})) \cdot {y_i}(v)}}{{{y_i}({G_i}({S_{\rm{A}}})) \cdot {y_i}(a)}} \notag\\ 
  & \Rightarrow \frac{{{y_i}({G_i}({S_{\rm{B}}} \cup v))}}{{{y_i}({G_i}({S_{\rm{B}}}))}} \ge \frac{{{y_i}({G_i}({S_{\rm{A}}} \cup v))}}{{{y_i}({G_i}({S_{\rm{A}}}))}} \notag\\ 
  & \Rightarrow {{\mathord{\buildrel{\lower3pt\hbox{$\scriptscriptstyle\frown$}} 
\over W} }_{{\rm{B}} \cup v - {\rm{A}} \cup v}} \ge {{\mathord{\buildrel{\lower3pt\hbox{$\scriptscriptstyle\frown$}} 
\over W} }_{{\rm{B - A}}}} 
\end{align}

Case (3). When $G_i (S_{\rm{B}} \cup v)$ executes Red-ope from $G_i (S_{\rm{B}} )$, $G_i (S_{\rm{A}} \cup v)$ will perform Add-ope, Exc-ope or Red-ope from $G_i (S_{\rm{A}} )$. In this case, $y_i (G_i (S_{\rm{B}} \cup v))= \frac{y_{i} (G_{i} (S_{\rm{B}} )) \cdot y_i (v)}{  \prod_{j}y_i (b_j)} $ where a class of components $b_j$ in $G_i (S_{\rm{B}} )$ is replaced by $v$. Since the cost-performance of $G_i (S_{\rm{A}} )$ must equal to or less than that of $G_i (S_{\rm{B}} )$, we can deduce similarly that $\prod_{j}y_i (b_j)$ will equal to or less than the value resulting from addition or replacements in $G_i (S_{\rm{A}} )$. Based on the analysis in Case (2), there is $\overset{\frown}{W}_{\rm{B} \cup v - \rm{A} \cup v} \geq \overset{\frown}{W}_{\rm{B}-\rm{A}}$.

As a result of three cases, it can be concluded that when $G_i (S_{\rm{A}} \cup v) \neq G_i (S_{\rm{A}} )$ and $G_i (S_{\rm{B}} \cup v) \neq G_i (S_{\rm{B}} )$, there is $\overset{\frown}{W}_{\rm{B} \cup v - \rm{A} \cup v} \geq \overset{\frown}{W}_{\rm{B}-\rm{A}}$, resulting similarly in $y_{i} (G_{i} (S_{\rm{A}} )) - y_i (G_i (S_{\rm{A}} \cup v)) \geq y_{i} (G_{i} (S_{\rm{B}} )) - y_i (G_i (S_{\rm{B}} \cup v))$.

In summary, there is the inequality $y_{i} (G_{i} (S_{\rm{A}} )) - y_i (G_i (S_{\rm{A}} \cup v)) \geq y_{i} (G_{i} (S_{\rm{B}} )) - y_i (G_i (S_{\rm{B}} \cup v))$ in three situations of TSSO with Markov property. Transforming $y_i (G_i (S))$ to $Y_i (G_i (S))$, it has $ Y_i (G_i (S_{\rm{A}} \cup v)) - Y_{i} (G_{i} (S_{\rm{A}} ))  \geq  Y_i (G_i (S_{\rm{B}} \cup v)) - Y_{i} (G_{i} (S_{\rm{B}} )) $, satisfying Definition 1, i.e., $Y_i (G_i (S))$ is submodular. According to TSSO formulation $\mathbb{E}[ Y_i (G_i (S))]$ and Lemma 1, it shows that the total objective function of TSSO is submodular when its sub-functions contain Markov and submodular features. \hfill $\square$\par

According to the Theorem A1, there are some observations.

\noindent \textbf{Observation A1.} In general, the non-submodularity of total objective function in TSSO derives from the mechanism that the non-additivity of marginal benefit is related with components that have been selected, rather than the scale of selective set.

\noindent \textbf{Proof:} Suppose that there are  subsets $S_{\rm{A}}$, $S_{\rm{B}}$, $v$ satisfying $S_{\rm{A}}  \subseteq S_{\rm{B}} \subseteq S_{\rm{L}}$, $v\in S_{\rm{L}} \backslash S_{\rm{B}}$, and $Y_i (G_i (S))$ is nondecreasing total function of TSSO. Imagining this situation that the benefit of component $v$ cannot be additive in $Y_{i} (G_{i} (S_{\rm{A}} ))$, i.e., $Y_i (G_i (S_{\rm{A}} \cup v)) = Y_{i} (G_{i} (S_{\rm{A}} ))$, due to the constraints. And there is $Y_i (G_i (S_{\rm{B}} \cup v)) > Y_{i} (G_{i} (S_{\rm{B}} ))$ since the previous set in $S_{\rm{B}}$ can recombine with $v$ to optimize the total objective value. Then there is $ Y_i (G_i (S_{\rm{A}} \cup v)) - Y_{i} (G_{i} (S_{\rm{A}} ))  <  Y_i (G_i (S_{\rm{B}} \cup v)) - Y_{i} (G_{i} (S_{\rm{B}} )) $ such that the total function of TSSO is not submodular. The reason for this situation is that the non-additivity in TSSO results in that increasing the feasible region in first stage cannot guarantee the expansion of effective feasible region in second stage. In other words, the expanding of feasible region in second stage is only connected to certain components rather than all unselected components. This situation is easily observed in some applications, like bipartite graph coverage \cite{balkanski2016learning}. \hfill $\square$\par

\noindent \textbf{Observation A2.} The sufficient conditions for total objective function in TSSO to be submodular are

\noindent 1) the sub-function is submodular, and

\noindent 2) there is no non-additivity in TSSO's sub-function, or its non-additivity is related to the selected set scale.

\ifCLASSOPTIONcaptionsoff
  \newpage
\fi



%



\bibliographystyle{IEEEtran}
\bibliography{ref}

%




\end{document}